\documentclass[%
 reprint,
 amsmath,amssymb,
 aps,
]{revtex4-2}
\usepackage [latin1]{inputenc}
\usepackage{graphicx}% Include figure files
\usepackage{dcolumn}% Align table columns on decimal point
\usepackage{bm}% bold math
\usepackage{hyperref}
\hypersetup{
    colorlinks=true,
    linkcolor=blue,
    filecolor=magenta,      
    urlcolor=blue,
}
\usepackage[]{physics}
\usepackage{tikz}
\usetikzlibrary{mindmap,trees}
\tikzset{every node/.append style={scale=0.9}}
\usetikzlibrary{quantikz}
\usepackage{caption}
\usepackage{subcaption}

\usepackage{multirow}

\begin{document}
\preprint{APS/123-QED}

\title{Molecular Excited State Calculations with \\
Adaptive Wavefunctions on a Quantum Eigensolver Emulation: \\
Reducing Circuit Depth and Separating Spin States}% Force line breaks with \\

\author{Hans Hon Sang Chan}
 \email{hans.chan@materials.ox.ac.uk}
\affiliation{%
 Department of Materials, University of Oxford, Parks Road, Oxford OX1 3PH, United Kingdom
}%

\author{Nathan Fitzpatrick}%
\affiliation{%
 Cambridge Quantum Computing Ltd., 9a Bridge Street, Cambridge CB2 1UB, United Kingdom
}%

\author{Javier Segarra-Mart\'i}
\affiliation{%
 Instituto de Ciencia Molecular, Universitat de Valencia, PO Box 22085 Valencia, Spain
 }%

\author{Michael J. Bearpark}%
\affiliation{%
 Department of Chemistry, Imperial College London, Molecular Sciences Research Hub, White City Campus, 82 Wood Lane, London W12 0BZ, United Kingdom
}%

\author{David P. Tew}%
\affiliation{%
 Physical and Theoretical Chemistry Laboratory, University of Oxford, South Parks Road, Oxford OX1 3QZ, United Kingdom
}%

\date{\today}% It is always \today, today,

\begin{abstract}
\textit{Ab initio} electronic excited state calculations are necessary for the quantitative study of photochemical reactions, but their accurate computation on classical computers is plagued by prohibitive resource scaling.
The Variational Quantum Deflation (VQD) is an extension of the quantum-classical Variational Quantum Eigensolver (VQE) algorithm for calculating electronic excited state energies, and has the potential to address some of these scaling challenges using quantum computers.
However, quantum computers available in the near term can only support a limited number of quantum circuit operations, so reducing the quantum computational cost in VQD methods is critical to their realisation.
In this work, we investigate the use of adaptive quantum circuit growth (ADAPT-VQE) in excited state VQD calculations, a strategy that has been successful previously in reducing the resource required for ground state energy VQE calculations.
We also invoke spin restrictions to separate the recovery of eigenstates with different spin symmetry to reduce the number of calculations and accumulation of errors in computing excited states.
We created a quantum eigensolver emulation package - Quantum Eigensolver Building on Achievements of Both quantum computing and quantum chemistry (QEBAB) - for testing the proposed adaptive procedure against two existing VQD methods that use fixed-length quantum circuits: UCCGSD-VQD and k-UpCCGSD-VQD.
For a lithium hydride test case we found that the spin-restricted adaptive growth variant of VQD uses the most compact circuits out of the tested methods by far, consistently recovers adequate electron correlation energy for different nuclear geometries and eigenstates while isolating the singlet and triplet manifold.
This work is a further step towards developing techniques which improve the efficiency of hybrid quantum algorithms for excited state quantum chemistry, opening up the possibility of exploiting real quantum computers for electronic excited state calculations sooner than previously anticipated.

% \begin{description}
% Secondary publications and information retrieval purposes.
% You may use the \texttt{description} environment to structure your abstract;
% use the optional argument of the \verb+\item+ command to give the category of each item. 
% \end{description}
\end{abstract}

\maketitle

%\tableofcontents

% \section{\label{sec:level1}First-level heading:\protect\\ The line
% break was forced \lowercase{via} \textbackslash\textbackslash}

\newpage
\section{Introduction}
Understanding how light interacts with molecules allows us to rationally harness it in many applications \cite{Gonzalez2020}.
Simulations of the underlying photochemical processes start with calculating not only ground but also excited electronic state energy profiles accurately. 
Conventional \textit{ab initio} methods adequate for this purpose however are computationally expensive \cite{Eriksen2020}.   
Thus, the calculation of electronic structure remains a major bottleneck in simulating excited state reactions, and has limited complete quantum treatment of light-matter interactions to small molecules.
This article investigates the use of adaptive quantum algorithms for the calculation of molecular excited state energies on emulated quantum computers, as a step towards addressing these ongoing challenges.

Motivated by quantum computers' ability to process exponentially more information than their classical counterparts, there has been much recent interest in using quantum computers as an alternative approach for  accurate electronic structure calculations, with some work on excited states \cite{McArdle2018, Cao2019, Bauer2020}.
Quantum algorithms generally encode molecular wavefunctions on \textit{qubits} and manipulate them using programmable operations (\textit{quantum gates}), then extract physically relevant expectation values by performing appropriate measurements of the qubits.
However, quantum computers available in the current Noisy Intermediate-Scale Quantum (NISQ) era are limited in their number of qubits, and the depth of \textit{quantum circuits} (sets of quantum gates) that can be applied before device noise corrupts their execution.

Variational Quantum Eigensolvers (VQE) are a class of quantum-classical hybrid algorithms that promises to bridge the gap between available NISQ computers and electronic structure calculations \cite{Cerezo2020a}.
These methods approximate molecular wavefunctions, or \textit{ans\"atze}, with parameterised, unitary operations $U(\vec{\theta})$:
\begin{equation}
    \ket*{\Psi(\vec{\theta})} = U(\vec{\theta})\ket{\psi_\text{ref}}
\end{equation}
where $\{\theta_i\}\in [0, 2\pi) $ and $\ket{\psi_\text{ref}}$ is an appropriate reference state, such as the mean-field solution to the electronic Schr\"odinger Equation.
An ansatz is then transformed into a parameterised quantum circuit, which encodes the unitary operations onto a register of qubits.
The quantum circuit expression for a Hermitian operator corresponding to an observable, such as the electronic Hamiltonian $\hat{H_{el}}$, is applied onto the ansatz, and the expectation value of its corresponding observable $E_{el}$ is extracted by repeated measurements of the qubits' values.
The ground state energy is obtained by variationally optimising the quantum circuit parameters until the energy converges:
\begin{equation}
    E_{el}(\vec{\theta}) = \min_{\vec{\theta}} \left( \bra*{\Psi(\vec{\theta})} \hat{H}_{el} \ket*{\Psi(\vec{\theta})} \right)
\end{equation}

The Unitary Coupled Cluster Singles and Doubles (UCCSD) is a popular, physically-motivated ansatz scheme for the VQE.
It is a variational and more robust alternative to the traditional coupled cluster method \cite{Lee2019, Evangelista2019}.
Although prohibitively inefficient on a classical computer \cite{Bartlett2007}, the parameterised excitation and relaxation operators in the Trotterised UCCSD ansatz can be mapped onto structured quantum circuit digits (\textit{Pauli gadgets}) which combine to prepare the wavefunction approximation efficiently on available qubits (see Supplementary Materials).

For excited state calculations, extension of the VQE in the Variational Quantum Deflation (VQD) method \cite{Higgott2019, Lee2019} is gaining traction, owing in part to the versatility of VQE in experimental demonstrations \cite{Peruzzo2014, Kandala2017, Colless2018a, Hempel2018, OMalley2016, Kokail2019, Ganzhorn2019, AIQuantum2020}, but also to VQD's simplicity and its ability to recover photochemical properties beyond just excited state energies\cite{Ibe2020}.
The VQD requires only a concurrent minimisation of the overlap between already computed eigenstates $\ket{\Phi_i}$ in addition to the variational energy expectation minimisation in VQE:
\begin{equation}
    F(\vec{\theta}) = \min_{\vec{\theta}} \left( \bra*{\Psi(\vec{\theta})} \hat{H} \ket*{\Psi(\vec{\theta})} + \sum_i \beta_i |\bra*{\Phi_i}\ket*{\Psi(\vec{\theta})}|^2 \right) \label{eq:vqd}
\end{equation}
where $\beta$ is a coefficient larger than the maximum energy gap that is to be calculated.

Although VQD is the leading candidate method for computing excited states on NISQ computers, there are a number of alternative quantum methods.
For example, a similar overlap measurement can be incorporated into the variational imaginary time evolution\cite{McArdle2018a, Endo2018} to find excited states.
The folded spectrum method\cite{Peruzzo2014} reformulates the VQE by variationally minimising the operator $H_\lambda = (H-\lambda I)^2$;
where the shift parameter $\lambda$ is an excited state eigenvalue of the Hamiltonian, the ansatz should converge to the corresponding eigenvector. 
The $H^2$ operator significantly increases the number of measurements needed, and finding a suitable shift parameter $\lambda$ which locates a desired eigenstate is difficult to control.
Quantum Subspace Expansion\cite{McClean2017} takes linear combinations of the converged ground state VQE ansatz $\sum_i O_i\ket{\Psi}$, where $O_i$ could correspond to fermionic excitation operators, to approximate a subspace of low-energy excited states.
The matrix elements of $\bra{\Psi}O_iHO_j\ket{\Psi}$ are calculated on a quantum computer, and the matrix is diagonalised on a classical computer.
However the quality of the excited state depends both on the quality of the ansatz and the linear expansion.
Finally, the Witness-Assisted Variational Eigenspectra Solver (WAVES)\cite{Santagati2018} minimises the ansatz energy expectation and the Von-Neumann entropy of a control ancilla qubit.
After the ground state has been found, the procedure starts a new VQE optimisation with an added entropy contribution to bias the ansatz towards an approximate excited state.
Tuning of the entropy contribution has similar shortfalls to QSE, but the main cost incurred here is the need for subsequent phase estimation, a very costly quantum subroutine, to extract the corresponding excited state energy from the approximate excited state ansatz.
Beyond using near-term noise-prone quantum computers, methods which use fault-tolerant quantum computers to recover excited state energies\cite{Bauman2019, Bauman2021} and energy gaps\cite{Sugisaki2021} have also been proposed. 

Most of the VQD schemes proposed hitherto use identical parameterised ansatz circuits $\ket*{\Psi(\vec{\theta})}$ for preparing both ground and excited states\cite{Higgott2019, Lee2019, Ibe2020}.
To ensure these fixed-length ans\"atze have sufficient variational flexibility for preparing different eigenstates, redundant parameterisation is often used.
In the UCC case, this is achieved by inclusion of generalised excitation and relaxation operators in the ans\"atze, such as UCC Generalised Singles and Doubles (UCCGSD) \cite{Higgott2019} and $k$-UCC Generalised Singles and paired Doubles ($k$-UpCCGSD) \cite{Lee2019}.
The increase in the number of operators leads to a commensurate increase of Pauli gadgets in the ansatz preparation circuit, rendering deep layers of simultaneous quantum gates (referred to as \textit{circuit depth}) which would be impossible to prepare noise-free on near-term quantum computers.
Estimation of the overlap between two eigenstates on a quantum computer further amplifies this problem:
overlap calculations on quantum computers use either the vacuum test, which doubles the circuit depth, or the destructive SWAP test, which doubles the number of qubits and has a large entangling controlled-NOT (CX) gate overhead \cite{Higgott2019} (entangling gates require qubits to interact, and are particularly prone to noise).
Thus, neither methods are NISQ-friendly, and they are made worse by the use of deep ansatz circuits. 
The large set of parameters $\{\theta_i\}$ in the generalised-excitation ans\"atze also burdens VQD with high classical optimisation overhead.
Finally, to find high energy excited states, many VQD studies \cite{Higgott2019, Ibe2020} required the sequential recovery of every eigenstate below the desired state.
As VQD is inherently prone to error propagation through each new energy calculation in the spectrum, this is not a desirable strategy.
The purpose of this work is to analyse how consolidating new VQE strategies and insights from quantum chemistry can reduce the quantum circuit depth for excited state preparation, and the number of states that need to have been calculated before reaching higher-energy states in VQD.

From work done by Greene-Diniz and Ramo, when spin restrictions are applied to the encoding of the starting mean-field reference state (herein referred to as a \textit{spin reference}), and also to the excitation operators included in the UCC ansatz, VQE can recover the lowest energy excited state which obeys the given symmetry \cite{Greene-Diniz2019}.
In the introduction of the $k$-UpCCGSD ansatz, Lee \textit{et. al.} considered the role of reference states in the calculation of the first excited electronic state of molecular nitrogen \cite{Lee2019}.
In their study of VQD using a hardware efficient ansatz to recover transition amplitudes, Ibe \textit{et. al.} also applied spin references other than the Hartree-Fock in their excited state preparation ansatz circuits \cite{Ibe2020}.
We use these concepts to systematically recover eigenstates of the electronic Hamiltonian confined to either the singlet or triplet manifold, using a combination of spin reference circuits and UCC-based spin-restricted ans\"atze.
This reduces the number of eigenstates that need to have been found first before reaching higher energy eigenstates.
Recovering eigenstates of a specific spin symmetry is a well-established feature in conventional classical electronic structure methods, such as Configuration Interaction (CI)\cite{Pancuz1979}.

Recent theoretical advances also gave rise to adaptive methods which iteratively grow the quantum circuit needed to capture the most important parts of the wavefunction for VQE calculations\cite{Grimsley2019, Tang2019, Rattew2019}.
In particular, the fermionic Adaptive Derivative-Assembled Pseudo-Trotter VQE (ADAPT-VQE), which iteratively selects excitation operators to append to an ansatz according to the system simulated, had higher accuracy and used less parameterised excitation operators than an equivalent ground state UCCSD-VQE calculation\cite{Claudino2020}.
This dramatically reduced the circuit depth for running ground state calculations to a given accuracy on a quantum computer, and will be needed to get close to running these calculations on real hardware for larger molecules in the near future.

ADAPT-VQE has been extended by Kottmann \textit{et. al.} to adaptively grow unique ans\"atze for each excited state in VQD calculations (ADAPT-VQD), combining quantum circuit efficiency with a physically motivated state preparation\cite{Kottmann2020a}.
%%%% Key summary %%%%%
% - testing it more systematically
% - using physical chemistry insight
% - document code that can be used
We complement their investigation by providing a systematic analysis of the circuit depth and accuracy of ADAPT-VQD in calculating excited state energies, by comparing it with the UCCGSD and $k$-UpCCGSD ans\"atze in VQD calculations.
We also highlight the use of spin restrictions at the reference level for isolated computations of singlet and triplet excited states.
We find that although adaptive ansatz growth has a larger measurement overhead than the other methods, the substantial reduction of circuit depth is an attractive feature for near-term quantum computing applications.
Furthermore, we show that VQD resources can be reduced by the separation of spin states.
We compare the VQD energies against exact Full Configuration Interaction (FCI) calculations across a range of geometries for the LiH molecule.
Finally, we document the Quantum Eigensolver Building on Achievements of Both (QEBAB) package that we created for testing these ideas \cite{chan_2021}.
%%%%%%

The article is organised as follows:
Section \ref{section:theory} provides the necessary background theory for UCC-type ans\"atze, the restriction of spin in state preparation, and adaptive spin-restricted VQD.
Section \ref{section:methods} contains computational details for the calculation of the LiH molecule's electronic structure in this investigation.
In Section \ref{section:results} we report and compare the different VQD emulations.
The Supplementary Material supplied with this work introduces the functionalities of the QEBAB code we created for this investigation, and also serves as a step-by-step introduction to VQE methods.

\section{Theory}
\label{section:theory}
\begin{figure*}
    \centering
    \resizebox{0.76\textwidth}{!}{%
    \begin{subfigure}[c]{0.5\textwidth}
        \centering
        \begin{quantikz}
        &&\lstick{$\ket{0}_0$} &  \gate{X} & \qw\\
        &&\lstick{$\ket{0}_1$} & \gate{X} & \qw\\
        &&\lstick{$\ket{0}_2$} & \qw & \qw\\
        &&\lstick{$\ket{0}_3$} & \qw & \qw
        \end{quantikz}\hspace{0.4cm}
        \includegraphics[scale=0.25]{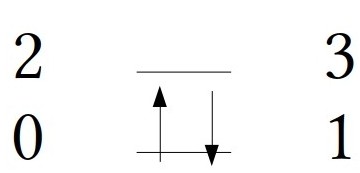}
        \begin{equation*}
            \ket{\psi_\text{HF}} = \ket{1100}
        \end{equation*}
    \end{subfigure}%
    \begin{subfigure}[c]{0.5\textwidth}
        \centering
        \begin{quantikz}
        &&\lstick{$\ket{0}_0$} &  \gate{X} & \qw\\
        &&\lstick{$\ket{0}_1$} & \qw & \qw\\
        &&\lstick{$\ket{0}_2$} & \gate{X} & \qw\\
        &&\lstick{$\ket{0}_3$} & \qw & \qw
        \end{quantikz}\hspace{0.4cm}
        \includegraphics[scale=0.25]{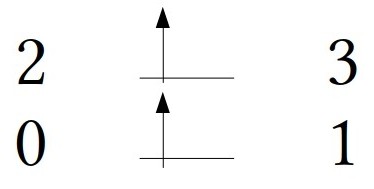}
        \begin{equation*}
            \ket{\psi_{T_1}} = \ket{1010}
        \end{equation*}
    \end{subfigure}
    }
    \caption{
    The closed-shell HF and the open-shell triplet reference circuit for a 4 spin-orbital system.
    }
    \label{fig:ref_circs}
\end{figure*}

\subsection{Unitary Coupled Cluster}
The UCC ansatz is given by Equation \ref{eq:UCC}:
\begin{equation}\label{eq:UCC}
    U(\vec{\theta}) = e^{\hat{T}(\vec{\theta}) - \hat{T}^\dagger(\vec{\theta})}
\end{equation} 
where the excitation operators are given by Equation \ref{eq:excite}:
\begin{eqnarray}\label{eq:excite}
    \hat{T}(\vec{\theta}) -\hat{T}^\dagger(\vec{\theta}) &=& \sum_{i,a} \theta_i^a (a_i^\dagger a_a - a_a^\dagger a_i) \nonumber \\ 
    & &+\sum_{i,j,a,b} \theta_{ij}^{ab} (a_i^\dagger a_j^\dagger a_aa_b - a_a^\dagger a_b^\dagger a_ia_j) \nonumber \\
    & &+ \dots
\end{eqnarray}
and $\{i,j\}$ indexes occupied orbitals and $\{a,b\}$ indexes unoccupied orbitals.
The above unitary $U(\vec{\theta})$ containing a single exponent is not executable on a quantum computer.
Instead, the Trotter formula\cite{Romero2019} is invoked to approximate the unitary:
\begin{equation}
    e^{A+B} = (e^Ae^B)^\rho 
\end{equation}
to approximate the unitary by taking a Trotter step $\rho=1$, sometimes referred to as the disentangled (factorised) UCC:
\begin{equation}
     U_{t1}(\vec{\theta}) = \prod_m e^{\theta_m(\tau_m - \tau_m^\dagger)}
\end{equation}
where $m$ indexes all possible excitations, $\theta_m \in \{\theta^a_i, \theta^{ab}_{ij}, \dots\}$ and $\tau_m \in \{a^\dagger_a a_i, a^\dagger_a a^\dagger_b a_ia_j\, \dots \}$.
It has been shown that specific orderings of operators in the disentangled UCC can exactly parameterise any state\cite{Evangelista2019}.

In practice, the excitations are truncated to only include single and double excitations.
This UCCSD ansatz has been popular in the VQE literature and is often the benchmark for more cost effective methods\cite{Romero2019}.
There are two variants of this ansatz that are suitable for VQD, both of which include generalised excitations in $U(\vec{\theta})=e^{\hat{T}(\vec{\theta}) - \hat{T}^\dagger(\vec{\theta})}$.
In the UCCGSD ansatz, the unitary includes the set of generalised single and double excitations\cite{Greene-Diniz2019}:
\begin{eqnarray}
    \hat{T}(\vec{\theta}) -\hat{T}^\dagger(\vec{\theta}) &=& \sum_{p,r} \theta_p^r (a_p^\dagger a_r - a_r^\dagger a_p) \nonumber \\
    & & + \sum_{p,q,r,s} \theta_{pq}^{rs} (a_p^\dagger a_q^\dagger a_ra_s - a_r^\dagger a_s^\dagger a_pa_q)
\end{eqnarray}
where $\{p,q,r,s\}$ indexes any orbital in the molecule, regardless of its occupancy.
In the $k$-UpCCGSD ansatz, the unitary is replaced by a sequence of generalised single excitations and only paired double excitations (pairwise excitations from one molecular orbital to another):
\begin{eqnarray}
    \hat{T}(\vec{\theta}) -\hat{T}^\dagger(\vec{\theta}) = \sum_{p,r} \theta_p^r (a_p^\dagger a_r - a_r^\dagger a_p) \nonumber \\
     + \sum_{p,r} \theta^{r_\alpha r_\beta}_{p_\alpha p_\beta} (a_{p_\alpha}^\dagger a_{p_\beta}^\dagger a_{r_\alpha} a_{r_\beta} - a_{r_\alpha}^\dagger a_{r_\beta}^\dagger a_{p_\alpha} a_{p_\beta})
\end{eqnarray}
and $k$ products of the unitary operator are taken for increased variational flexibility:
\begin{equation}
    U(\vec{\theta}_1, \vec{\theta}_2, ... ,\vec{\theta}_k) = e^{\hat{T}(\vec{\theta}_1) - \hat{T}^\dagger(\vec{\theta}_1)}e^{\hat{T}(\vec{\theta}_2) - \hat{T}^\dagger(\vec{\theta}_2)}...e^{\hat{T}(\vec{\theta}_k) - \hat{T}^\dagger(\vec{\theta}_k)}
\end{equation}
The sequence of unitary operators are then translated into quantum circuit logic gates, using a number of different available methods.
In this work the Jordan-Wigner transform was used, which transforms the unitary operators into parameterised tensor products of Pauli operators (\textit{Pauli strings}), and encodes the occupancy of a spin orbital onto the binary value of each qubit.
The parameterised quantum circuit form of these operators can be found in the Supplementary Material.

\subsection{Spin-Restriction in State Preparation}
Spin symmetry restrictions confine the wavefunction Hilbert space explored by the ansatz to a smaller subspace with the desired eigenstate.
In this work, state preparation ans\"atze were spin-restricted at the reference and parameterised unitary level.
The closed-shell Hartree-Fock reference state $\ket{\psi_\text{HF}}$ was used for calculating singlet states, and the open-shell triplet reference $\ket{\psi_{T_1}}$ for triplet states (see Figure \ref{fig:ref_circs}).

Only UCCSD excitation operators which conserve spin %(i.e. only $\alpha\rightarrow\alpha$ and $\beta\rightarrow\beta$ excitations)
were used for ansatz construction.
Ans\"atze generated in this manner preserve the spin symmetry of the spin reference:
\begin{align}
    \ket{\Psi^S(\vec{\theta})} =& \prod_m e^{\theta_m\hat{A}_m^S}\ket{\psi_\text{HF}}\\
    \ket{\Psi^T(\vec{\theta})} =& \prod_m e^{\theta_m\hat{A}_m^S}\ket{\psi_{T_1}}
\end{align}
where $ \hat{A}^S_m \in  \{ E_i^a - E_a^i, E_{ij}^{ab} - E_{ab}^{ij} \} $ are \textit{spin-adapted} operators which consist of unitary group generators
($E_i^a = a_{i\alpha}^\dagger a_{a\alpha} + a_{i\beta}^\dagger a_{a\beta}$ and
$E_{ij}^{ab} = a_{i\alpha}^\dagger a_{j\alpha}^\dagger a_{a\alpha}a_{b\alpha} + a_{i\beta}^\dagger a_{j\beta}^\dagger a_{a\beta}a_{b\beta} + a_{i\alpha}^\dagger a_{j\alpha}^\dagger a_{a\beta}a_{b\beta} + a_{i\beta}^\dagger a_{j\beta}^\dagger a_{a\alpha}a_{b\alpha}$
are singlet single and double excitations respectively,
where $i,j,a,b$ are indices of molecular spatial orbitals).
This scheme can in principle be applied to any UCC-based ansatz.

In practice, not every formulation of spin-adapted UCC ansatz can be encoded onto a quantum computer.
In particular, the Pauli strings resulting from some of the spin-adapted double excitations $E_{ij}^{ab} - E_{ab}^{ij}$ do not commute, and the spin adaptation is broken.
For example, some UCCGSD excitations are expected to not observe spin-restrictions when encoded into a quantum circuit form, whereas $k$-UpCCGSD excitations do.

\begin{figure*}
    \includegraphics[scale=0.35]{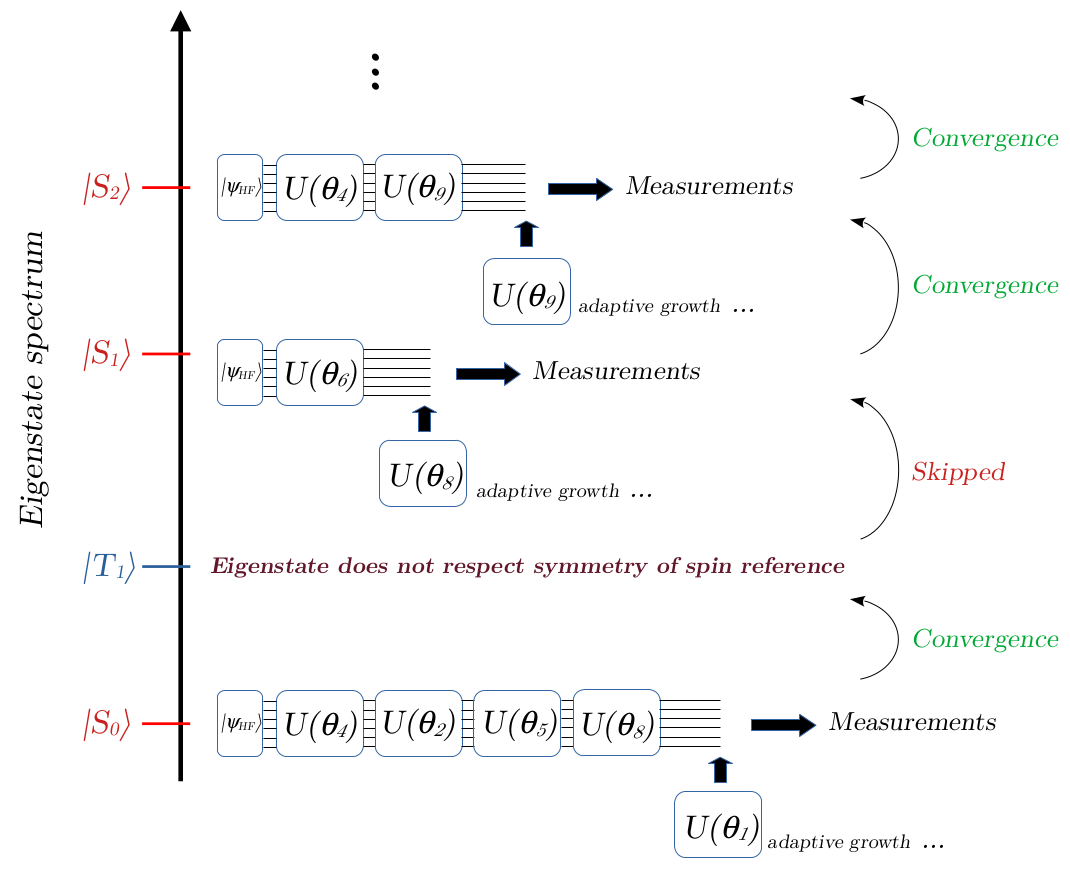}
    \caption{Diagrammatic scheme of ADAPT-VQD using the singlet spin reference.
    Where an eigenstate does not respect the symmetry of the spin reference, the algorithm simply converges to the next eigenstate that does.
    This method can potentially ensure that the quantum circuit at each eigenstate is as compact as possible and uniquely recovers the corresponding correlation energy.}
\end{figure*}

\subsection{Spin-Restricted ADAPT-VQD}
ADAPT-VQE\cite{Grimsley2019, Tang2019} iteratively grows the ansatz circuit by selecting operators $\hat{A}_m$ which are expected to recover the most significant amount of electron correlation from a predefined pool
(e.g. all spin-adapted single and double excitations $ \hat{A}^S_m \in  \{ E_i^a - E_a^i, E_{ij}^{ab} - E_{ab}^{ij} \}$
) and sequentially appending them onto the ansatz.
At each new iteration, the energy is variationally optimised with the new operator included in the ansatz, and the ansatz growth is terminated when the energy converges to a defined threshold.
It has been shown that at the infinite limit this ansatz is expected to converge to the exact wavefunction\cite{Grimsley2019}:
% {\color{red} DPT: hmm - the number of parameters in the FCI wavefunction is not infinite!}
\begin{equation}
    \ket*{\Psi_\text{FCI}} = \prod^\infty_k \prod^{N^4}_{m} e^{\theta^k_m \hat{A}_m}\ket{\psi_\text{HF}} 
\end{equation}
The candidate operator $\hat{A}_m$ from the pool with the largest absolute energy gradient (which, when added, influences the energy the most) was the proposed selection criterion at each iteration:
\begin{equation}
    \pdv{E}{\theta_m} = 2\mathcal{R} \bra{\Psi(\vec{\theta})}\hat{A}_m\hat{H}\ket{\Psi(\vec{\theta})} \label{eq:gradient}
\end{equation}
The quantum circuit involved for this measurement on quantum computers\cite{McClean2017a} can be found in the \hyperref[supplementary]{Supplementary Material}.
The convergence threshold was defined by the norm $\epsilon$ of the vector composed of all energy gradient computations:
\begin{equation}
    \epsilon = \sqrt{\sum_m \left(\pdv{E}{\theta_m}\right)^2 }
\end{equation}

Previous simulations\cite{Grimsley2019} of molecular ground states showed that this addresses the main limitations of the UCCSD-VQE scheme:
firstly, the ordering of the operators in the ansatz wavefunction is well defined.
The measured energy expectation can thus be kept invariant between different implementations of ADAPT-VQE.
ADAPT-VQE is also competitive in performance against the UCCSD ansatz when it comes to recovering sufficient correlation energy, despite using less excitation operators.
Although this method necessitates a gradient measurement overhead for every candidate operator at every iteration, the significantly shorter circuit means that accurate recovery of ground state energies may be possible even on relatively noisy quantum computers.

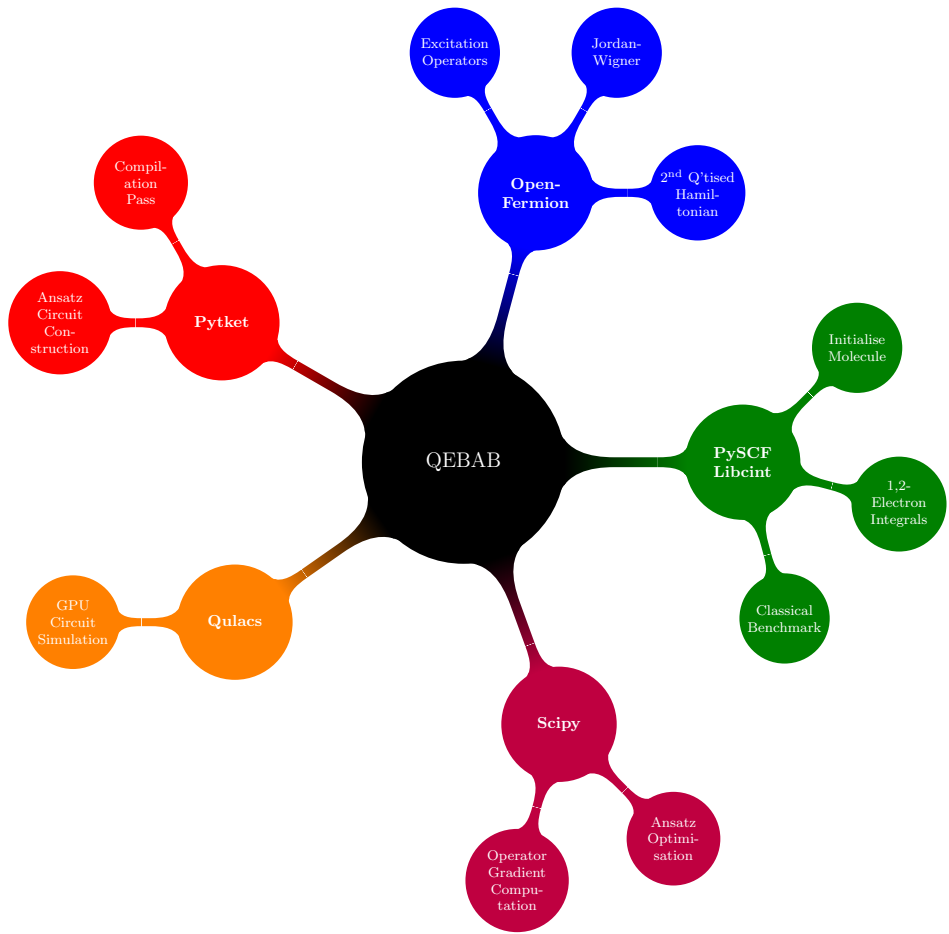
\begin{figure*}
    \centering
    \resizebox{.7\textwidth}{!}{%
        \begin{tikzpicture}
        \def\nchilds{5}
        \path[mindmap,concept color=black,text=white]
        node[concept] {QEBAB}
        [clockwise from=0]
        child[concept color=green!50!black] { node[concept] {\bf{PySCF Libcint}}
            [clockwise from=45]
            child { node[concept] {Initialise Molecule} }
            child { node[concept] {1,2-Electron Integrals} }
            child { node[concept] {Classical Benchmark} }
        }
        child[concept color=blue, grow=75] { node[concept] {\bf{Open-Fermion}}
            [clockwise from=120]
            child { node[concept] {Excitation Operators} }
            child { node[concept] {Jordan-Wigner} }
            child { node[concept] {2$^\text{nd}$ Q'tised Hamiltonian} }
        }
        child[concept color=red, grow=150] { node[concept] {\bf{Pytket}}
            [clockwise from=180]
            child { node[concept] {Ansatz Circuit Construction} }
            child { node[concept] {Compil-ation Pass} }
        } 
        child[concept color=orange, grow=215] { node[concept] {\bf{Qulacs}}
            [clockwise from=180]
            child { node[concept] {GPU Circuit Simulation} }
        }
        child[concept color=purple, grow=290] { node[concept] {\bf{Scipy}}
            [clockwise from=-45]
            child { node[concept] {Ansatz Optimisation} }
            child { node[concept] {Operator Gradient Computation} }
        };
        \end{tikzpicture}
    }
    \caption{We introduce the Quantum Eigensolver Building on Achievments of Both quantum computing and quantum chemistry package, which we developed for this investigation.
    This is a summary of the libraries used in the code.}
    \label{fig:code}
\end{figure*}

We extend the ADAPT-VQE to express excited states by including orthogonal constraints in the operator selection and energy optimisation, negating the need for fixed-length, generalised ans\"atze that map to long circuits.  
In VQD, the effective Hamiltonian of the $j^\text{th}$ eigenstate with the orthogonal penalisation is:
\begin{equation}
    \hat{H}_j = \hat{H} + \sum^{j-1}_{i=0} \beta_i \ket{\Phi_i}\bra{\Phi_i}
\end{equation}
where $\ket{\Phi_i}$ is the $i^\text{th}$ previously optimised lower energy state.
To make VQD adaptive, the effective Hamiltonian is substituted into the energy gradient with respect to candidate operators (Equation \ref{eq:gradient}):
\begin{eqnarray}
    \pdv{E}{\theta_m} &=& 2\mathcal{R} \bra*{\Psi(\vec{\theta})}\hat{A}^S_m\hat{H}\ket*{\Psi(\vec{\theta})} \nonumber \\
    & & + 2\mathcal{R}\sum^{j-1}_{i=0} \beta_i\bra*{\Psi(\vec{\theta})}\hat{A}^S_m \ket*{\Phi_i}\bra*{\Phi_i}\ket*{\Psi(\vec{\theta})}
    \label{eq:ov_grad}
\end{eqnarray}
where $\ket*{\Psi(\vec{\theta})}$ is the currently sought state, and $\hat{A}^S_m \in  \{ E_i^a - E_a^i, E_{ij}^{ab} - E_{ab}^{ij} \}$ is a candidate spin-adapted excitation operator in the pool.
The energy gradient is thus separable into two parts; the energy gradient component used in the original ADAPT-VQE, and an overlap gradient component.
Both components can be measured in parallel across multiple quantum processors for each candidate operator in the pool.
The derivation of Equation \ref{eq:ov_grad} and the corresponding circuit can also be found in the \hyperref[supplementary]{Supplementary Material}.

At each eigenstate, our spin restricted ADAPT-VQD uses the energy gradient defined above to iteratively select operators with the largest energy gradients from a sufficiently expressive operator pool to grow the state ansatz.
When the energy is considered converged at a given eigenstate, the ansatz is stored and the procedure moves on to seek the next orthogonal eigenstate.
As correlation energy is a function of eigenstate, higher energy ans\"{a}tze are expected to use different operators.

\section{Computational Methods}
\label{section:methods}
We developed a Python package, Quantum Eigensolver Building on Achievements of Both quantum computing and chemistry (QEBAB), to construct and emulate the spin-restricted UCCGSD, $k$-UpCCGSD and aforementioned ADAPT ansatz circuit variants for computing the lowest-lying excited state energies of a target molecule.
Figure \ref{fig:code} provides a summary of the code's features, which is interfaced with PySCF\cite{Sun2018} and Libcint \cite{Libcint} for extracting the required one- and two-electron integrals and initialising the molecule, Open-Fermion\cite{McClean2020} for creating the second quantised Hamiltonian and transformation of UCC excitation operators into Pauli gadgets, Pytket\cite{Sokolov2020} for construction and compilation of ansatz circuits, and Qulacs \cite{suzuki2020qulacs} for GPU-accelerated circuit simulation.
Additionally, Scipy\cite{Virtanen2020} was employed for variational minimisation of ansatz expectation energies.
Details of the code can be found in the Supplementary Material.

We note here that although the Pytket library has the capacity to compile circuits for NISQ computers that are available today, the state of the art technology is too noisy for demonstrating the proof-of-principle in this work.
We therefore only emulated the quantum calculations on conventional computers, which has an exponential slow down because simulation of an elementary quantum logic gate requires the update of all $2^\mathcal{N}$ coefficients in the qubit statevector ($\mathcal{N}$ is the number of qubits).
When more robust quantum hardware becomes available, these quantum methods are expected to be competitive against classical quantum chemistry computation.

\begin{figure}[h!]
    \centering
        \begin{subfigure}[c]{0.45\textwidth}
            \centering
            \includegraphics[scale=0.4]{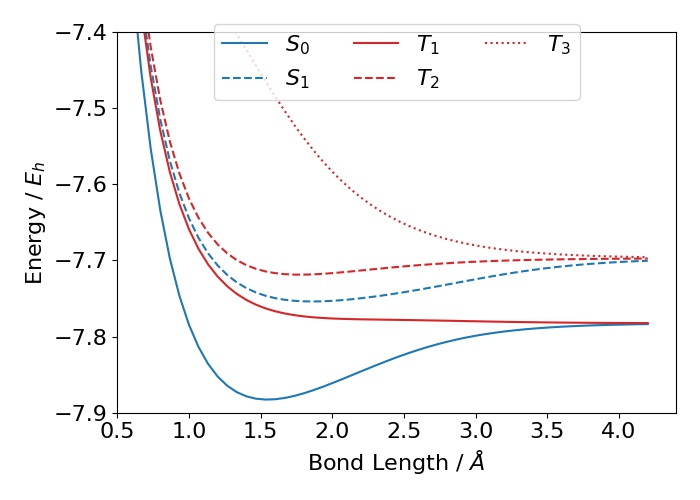}    
        \end{subfigure}
        \begin{subfigure}[c]{0.45\textwidth}
            \centering
            \begin{tabular}{c|cc}
            \toprule
            State & Irr. Rep. & Excitation \\
            \colrule
            $T_3$ & $A_1$ & $\sigma \rightarrow p_z $\\
            $T_2$ & $E_{1x}, E_{1y}$ & $\sigma \rightarrow p_x,p_y$\\
            $S_1$ & $A_1$ & $\sigma \rightarrow \sigma^\ast$\\
            $T_1$ & $A_1$ & $\sigma \rightarrow \sigma^\ast$\\
            $S_0$ & $A_1$ & - \\ 
            \end{tabular}
        \end{subfigure}%
    \caption{The FCI electronic structure calculation of the lowest-lying eigenstates along the Li-H bond length.
    Red curves indicate triplet states and blue curves singlet states. }
    \label{fig:FCI}
\end{figure}

In this work, the LiH molecule with the minimal STO-3G basis was chosen for comparing the performance of different approaches.
As the number of qubits needed corresponds to the number of spin orbitals, the exponential classical emulation cost of $\mathcal{N}$ qubits limits the test cases reported herein to such a small, unphysical one-electron basis set.
The number of qubits available on near-term NISQ machines is also only on the order of $10^2$,\cite{Arute2019} so it is conceivable that the first experimental demonstration of the methods related to this work will also be limited to small systems and basis sets.
The Full Configuration Interaction (FCI) method, as implemented in PySCF\cite{Sun2018}, was used to calculate the energies of the lowest-lying eigenstates for Li-H bond lengths of 0.40\AA \ and 4.20\AA \ at 0.07\AA \ intervals (see Figure \ref{fig:FCI}).
The ordering of the energy states were in agreement with \cite{Docken1972}, although the values were different because a different basis set was used.
From the FCI calculation, we chose only a small number of representative geometries to benchmark the VQD methods due to the aforementioned emulation cost consideration.

For the UCCGSD and spin-restricted $k$-UpCCGSD, calculations of the lowest two singlet and triplet eigenstates were simulated, at bond lengths of 0.91\AA, 1.00\AA, 1.80\AA, 2.02\AA \ and 2.98\AA.
For the spin-restricted ADAPT-VQD, we could afford to calculate eigenstates using ans\"atze grown from both the singlet and triplet spin references respectively, at nine bond lengths between 1.00\AA \ and 4.2\AA \ at 0.40\AA \ intervals.
This system at STO-3G is described with 6 $\alpha$- and 6 $\beta$-spin orbitals, so 12-qubit circuits were simulated for all cases.
At this minimal basis, the emulated UCC-type ans\"atze are expected to yield energies close to the FCI.
The real-valued coefficient $\beta$ in the orthogonality penalisation term should be sufficiently larger than the greatest energy gap recovered\cite{Higgott2019} without incurring a high optimisation cost for excited states.
Noting the maximum energy differences from the FCI calculation, a compromise was set to $\beta=3.0$ for all VQD calculations.
For ADAPT-VQD, the operator norm convergence threshold was set to be $\epsilon=0.01$, which was deemed adequate for small molecules in previous studies\cite{Grimsley2019, Claudino2020}.
The pool of spin-adapted UpCCGSD excitation operators was used for the ADAPT procedure, where operators are not exhausted and the same operator can in principle be added to an ansatz multiple times.
The Jordan-Wigner (JW) encoding was used to map the unitary fermionic excitation operators into Pauli gadget circuits (refer to Supplementary Material for details), and Pytket's compilation tools\cite{Cowtan2019, cowtan2020generic} were used to reduce the circuit depth in every ansatz construction.
L-BFGS-B was the optimiser of choice for the variational minimisation of the ansatz parameters.

\begin{figure*}
        \centering
        \begin{subfigure}[c]{\textwidth}
            \centering
            \caption{}
            \includegraphics[scale=0.35]{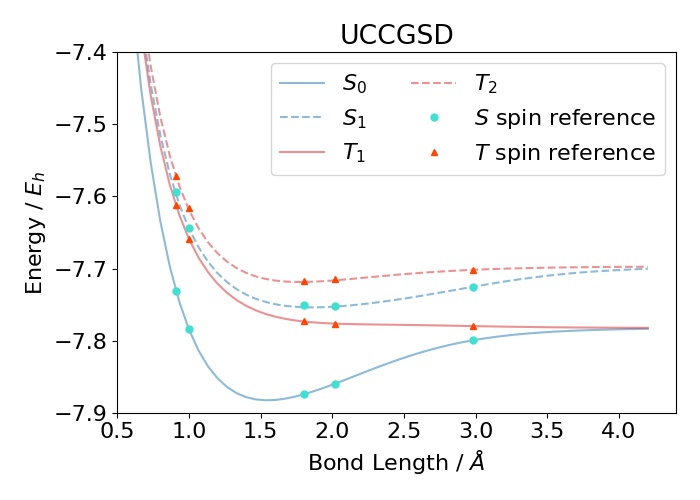}   
        \end{subfigure}
        \begin{subfigure}[c]{0.45\textwidth}
            \centering
            \includegraphics[scale=0.35]{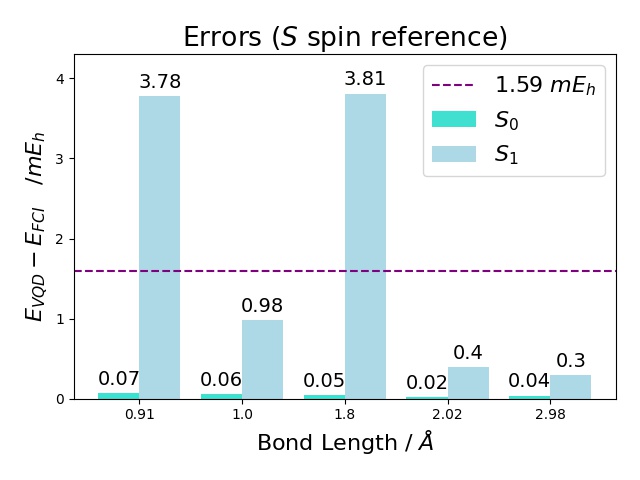}
            \caption{}
        \end{subfigure}%
        \begin{subfigure}[c]{0.45\textwidth}
            \centering
            \includegraphics[scale=0.35]{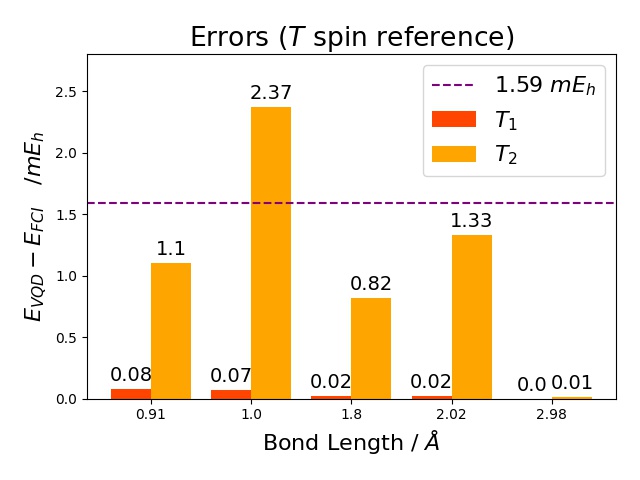}   
            \caption{}
        \end{subfigure}
        \caption{Quantum simulations of the singlet and triplet electronic excited states of LiH along the Li-H bond length using the UCCGSD ansatz.
        (A) The potential energy profile (in Hartree), where solid lines indicate FCI calculations and the markers refer to simulated UCCGSD values.
        Energy errors (in milliHartree) of (B) singlet and (C) triplet VQD energy calculations.}
        \label{fig:UCCGSD}
\end{figure*}
\begin{table*}
    \centering
    \resizebox{.9\textwidth}{!}{%
    \begin{tabular}{l|cc|cc|cc|cc|cc|c}
    \multirow{2}{*}{States}&\multicolumn{2}{c}{0.91}&\multicolumn{2}{c}{1.00}&\multicolumn{2}{c}{1.80}&\multicolumn{2}{c}{2.02}&\multicolumn{2}{c|}{2.98}&\multirow{2}{*}{NPE /$mE_h$}\\
                           & FCI & VQD                 & FCI & VQD                & FCI & VQD               & FCI & VQD               & FCI & VQD                & \\ 
    \hline
    $T_2$                  & -7.57227 & -7.57117       & -7.61856 & -7.61619      & -7.71859 & -7.71777     & -7.71629 & -7.71496      & -7.70177 & -7.70176  & 2.36 \\
    $S_1$                  & -7.59780 & -7.59401       & -7.64450 & -7.64352      & -7.75366 & -7.74986     & -7.75275 & -7.75236      & -7.72531 & -7.72501  & 3.51 \\
    $T_1$                  & -7.61242 & -7.61233       & -7.65893 & 7.65886       & -7.77343 & -7.77342     & -7.77637 & -7.77635      & -7.77980 & -7.77980  & 0.75 \\
    $S_0$                  & -7.73064 & -7.73057       & -7.78446 & -7.78440      & -7.87452 & -7.87448     & -7.85959 & -7.85956      & -7.79952 & -7.79948  & 0.05 \\
    \colrule
    \end{tabular}
    }
    \caption{The values of the UCCGSD-VQD energies at the computed bond lengths (in Hartree), and the NPE of recovered eigenstates (in milliHartree).}
    \label{tab:UCCGSD}
\end{table*}

To verify that the final optimised ans\"atze obey the applied spin restrictions, their $S^2$ spin expectation values were calculated in the same manner as the energy expectation calculations (given a second quantised expression of the $S^2$ operator), details in the Supplementary Material.
We use the nonparallelity error (NPE) to characterise the consistency of each computed eigenstate energy, a metric often employed in the variational quantum algorithm literature for comparing performance of different ans\"atze\cite{Romero2019, Lee2019, Sokolov2020}.
The NPE is the difference between the maximum and minimum error relative to FCI across all computed geometries of a given eigenstate:
\begin{equation}
    \text{NPE} = \text{max}(E_{\text{VQE}} - E_{\text{FCI}}) - \text{min}(E_{\text{VQE}} - E_{\text{FCI}})
\end{equation}

\section{Results}
\label{section:results}
\begin{figure*}
    \centering
    \begin{subfigure}[c]{\textwidth}
        \centering
        \caption{}
        \includegraphics[scale=0.4]{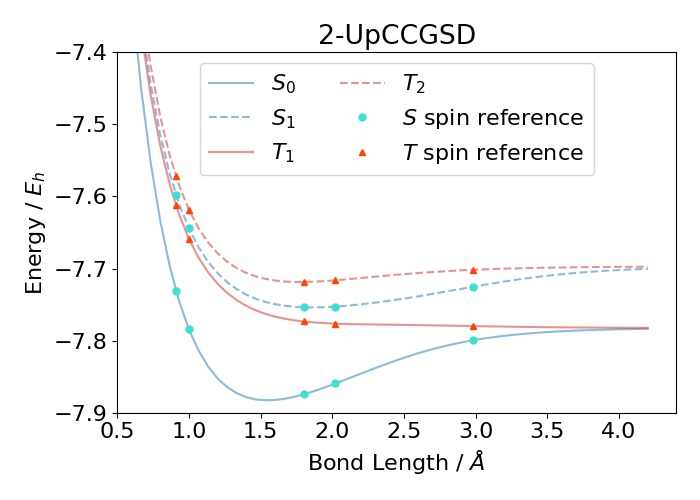}    
    \end{subfigure}
    \begin{subfigure}[c]{0.45\textwidth}
        \centering
        \includegraphics[scale=0.40]{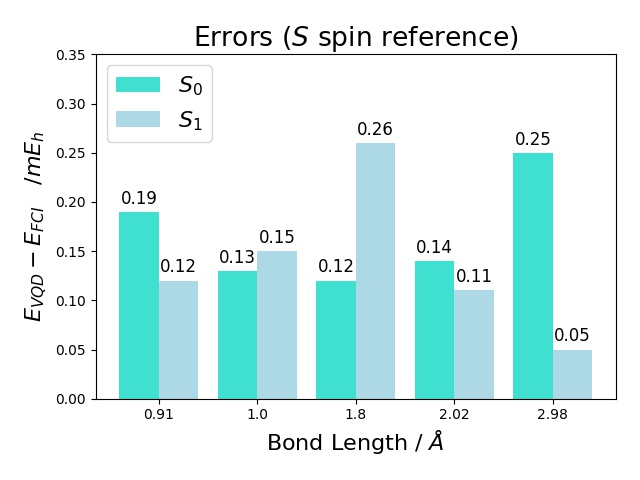}
        \caption{}
    \end{subfigure}%
    \begin{subfigure}[c]{0.45\textwidth}
        \centering
        \includegraphics[scale=0.40]{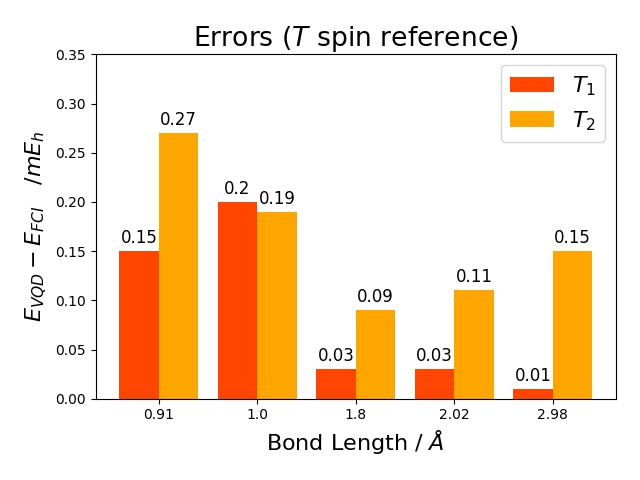}
        \caption{}
    \end{subfigure}
    \caption{Quantum simulations of the singlet and triplet electronic excited states of LiH along the Li-H bond length using the 2-UpCCGSD ansatz.
    (A) The potential energy profile (in Hartree), where solid lines indicate FCI calculations and the markers refer to simulated 2-UpCCGSD values.
    Energy errors (in milliHartree) of (B) singlet and (C) triplet VQD energy calculations.
    \label{fig:kUpCCGSD}}
\end{figure*}
\begin{table*}
    \centering
    \resizebox{.9\textwidth}{!}{%
    \begin{tabular}{l|cc|cc|cc|cc|cc|c}
    \multirow{2}{*}{States}&\multicolumn{2}{c}{0.91}&\multicolumn{2}{c}{1.00}&\multicolumn{2}{c}{1.80}&\multicolumn{2}{c}{2.02}&\multicolumn{2}{c|}{2.98}&\multirow{2}{*}{NPE /$mE_h$}  \\
                           & FCI & VQD                 & FCI & VQD                & FCI & VQD               & FCI & VQD               & FCI & VQD                & \\ 
    \hline
    $T_2$                  & -7.57227 & -7.57200       & -7.61856 & -7.61837      & -7.71859 & -7.71850     & -7.71629 & -7.71618     & -7.70177 & -7.70162      & 0.18 \\
    $S_1$                  & -7.59780 & -7.59767       & -7.64450 & -7.64435      & -7.75366 & -7.75341     & -7.75275 & -7.75264     & -7.72531 & -7.72526      & 0.21 \\
    $T_1$                  & -7.61242 & -7.61227       & -7.65893 & -7.65873      & -7.77343 & -7.77340     & -7.77637 & -7.77634    & -7.77980 & -7.77979      & 0.19 \\
    $S_0$                  & -7.73064 & -7.73044       & -7.78446 & -7.78433      & -7.87452 & -7.87440     & -7.85959 & -7.85945     & -7.79952 & -7.79927      & 0.11 \\
    \colrule
    \end{tabular}
    }
    \caption{The values of the 2-UpCCGSD-VQD energies at the computed bond lengths (in Hartree), and the NPE of recovered eigenstates (in milliHartree).}
    \label{tb:2UpCCGSD}
\end{table*}

\subsection{UCCGSD-VQD} \label{section:uccgsd-vqd}
    The UCCGSD ansatz for LiH had 135 independent spin-adapted excitation operators.
    This mapped to 1860 Pauli gadget circuit units under the JW transform, which totalled to, after compilation pass, circuit and CX-gate depths of 6368 and 3904 gates to encode.
    The simulations showed that the singlet spin reference imposed was indeed successful in exclusively calculating only to the $S_0$ and $S_1$ states, avoiding the $T_1$ state.
    Likewise, when the triplet spin reference was used in conjunction with the ansatz containing only singlet excitations, it also optimised to the $T_1$ and doubly degenerate $T_2$ states.
    The ground state energies $S_0$ and $T_1$ were very accurate as expected\cite{Lee2019}, with NPE of 0.05 m$E_h$ and 0.07 m$E_h$ (Table \ref{tab:UCCGSD}).
    However, the accuracy was poor for the excited states $S_1$ and $T_2$; the error plots in Figure \ref{fig:UCCGSD} show that many of the calculated energies failed to reach within 1.59 m$E_h$ (chemical accuracy) of the exact energy.
    This poor performance can be attributed to the large parameterisation landscape from the excitation operators used in the ansatz, which is likely to give incomplete optimisations towards local minima.
    Use of alternative optimisation algorithms can mitigate this issue, though they might come at a cost of more energy expectation measurements. 
    Furthermore, as mentioned in Section \ref{section:theory}, the quantum circuit representation of this ansatz is expected to break spin-adaptation even if spin-adapted fermionic excitations were used.
    As such, the average $S^2$ expectation value for the optimised $S_1$ state UCCGSD ans\"atze was 0.08.
    However, the $S^2$ expectation for the optimised $S_1$, $T_1$ and $T_2$ ans\"atze were 0.00, 2.00 and 2.00 across all bond lengths, suggesting that it is possible to obtain spin-pure states but it is not guaranteed.
    Nonetheless, coupled with the prohibitively long ansatz circuits, the unstable spin control over and poor accuracy of the UCCGSD ansatz renders it an impractical choice for computing excited states on NISQ-era devices.

\subsection{Spin-Restricted $k$-UpCCGSD-VQD}
    In this investigation the 1-UpCCGSD and 2-UpCCGSD ans\"atze were tested.
    The 1-UpCCGSD ansatz has 30 excitation operators, corresponding to 180 Pauli gadget circuit units and a total 470-gate deep state preparation circuit.
    The simulations were not able to recover any of the excited eigenstates, likely due to the reduced expressiveness of the ansatz.
    
    The ansatz at $k=2$ doubled the resources at 60 excitation operators, 360 Pauli gadget circuit units, and a circuit depth of 941 gates and CX gate depths of 580.
    Despite the use of duplicated operators, the omission of unpaired double electron excitations saw an 80\% circuit resource reduction over the UCCGSD ansatz.
    Figure \ref{fig:kUpCCGSD} shows that the errors for the singlet and triplet ground state calculations were on average a factor of 2 higher than those calculated with the UCCGSD ansatz, but are still comfortably within chemical accuracy.
    Apart from the reduced circuit depth, another major improvement over the previous technique were the calculations of the first singlet and triplet excited states; the errors were within chemical accuracy consistently across all computed bond lengths at both $S_1$ and $T_2$, reporting NPE of 0.21 and 0.18 respectively (Table \ref{tb:2UpCCGSD}).
    The $S^2$ values were 0.00 for all singlet states and 2.00 for all triplet states, confirming successful application of spin-restriction.  
    
    On average, the number of optimisation steps for reaching convergence was 100 for UCCGSD-VQD calculations, but for 2-UpCCGSD-VQD the number of steps were doubled even though the latter has only half as many parameters.
    On a real quantum device, this procedure will accordingly require double the number of measurements.

\begin{figure*}
        \centering
        \begin{subfigure}[c]{\textwidth}
            \centering
            \caption{}
            \includegraphics[scale=0.35]{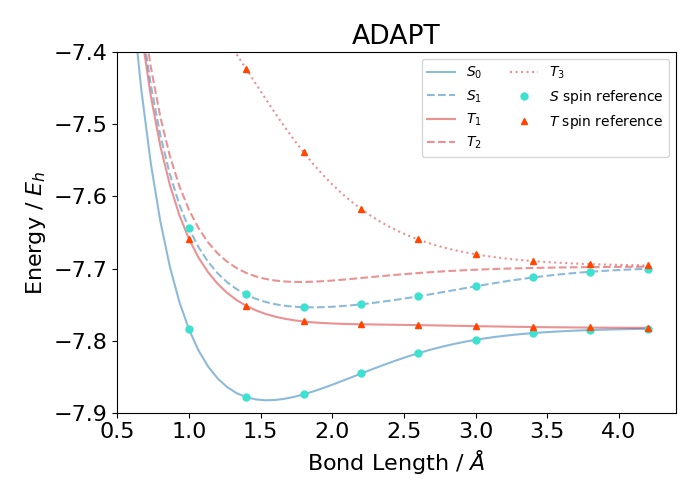}    
        \end{subfigure}
        \begin{subfigure}[c]{0.45\textwidth}
            \centering
            \includegraphics[scale=0.39]{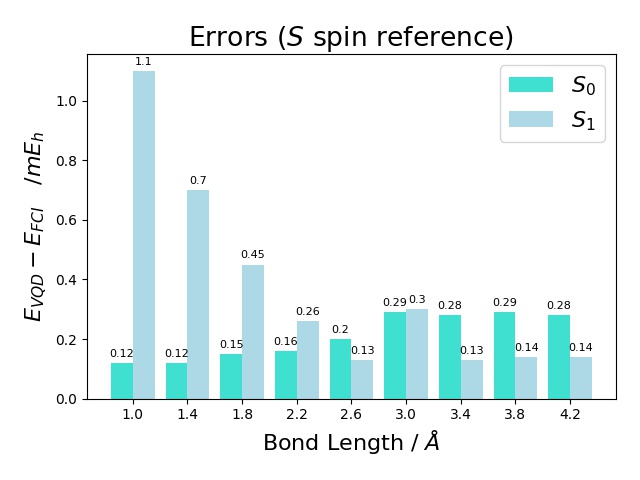}
            \caption{}
        \end{subfigure}%
        \begin{subfigure}[c]{0.45\textwidth}
            \centering
            \includegraphics[scale=0.39]{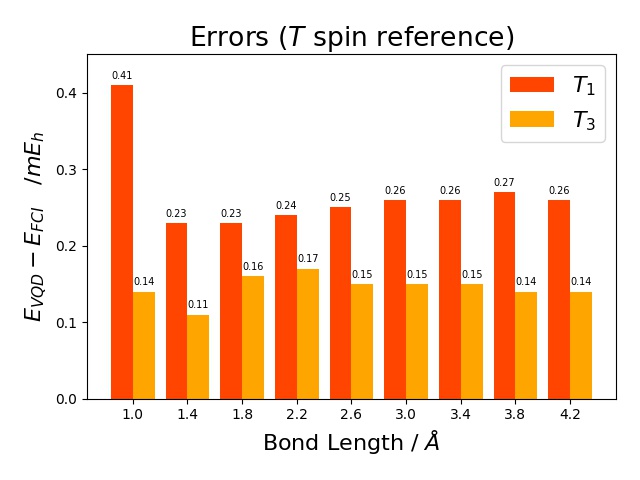}
            \caption{}
        \end{subfigure}
        \caption{Quantum simulations singlet and triplet excited states of \textnormal{LiH} at 5 bond lengths using adaptively grown ans\"atze, unique for each geometry and each eigenstate.
        (A) The potential energy profile (in Hartree), where solid lines indicate FCI calculations and the markers refer to simulated values.
        Although the singlet spin restriction recovered both the lowest lying singlet states, the triplet spin reference only recovered the $T_1$ and $T_3$ states, missing the doubly degenerate $T_2$ state.
        BOTTOM: The errors associated with each VQD energy calculation.
        Energy errors (in milliHartree) of (B) singlet and (C) triplet VQD energy calculations.}
        \label{fig:ADAPT}
    \end{figure*}%
    \begin{table*}
    \centering
    \resizebox{.9\textwidth}{!}{%
    \begin{tabular}{l|cc|cc|cc|cc|cc|}
    \multirow{2}{*}{States}&\multicolumn{2}{c}{1.00}& \multicolumn{2}{c}{1.40}& \multicolumn{2}{c}{1.80}&\multicolumn{2}{c}{2.20}&\multicolumn{2}{c|}{2.60}  \\
                           & FCI & VQD                 & FCI & VQD                & FCI & VQD               & FCI & VQD               & FCI & VQD             \\ 
    \hline
    $T_3$                  & -7.29798 & -7.29784       & -7.42335 & -7.42323      & -7.53882 & -7.53866     & -7.61737 & -7.61720     & -7.65974 & -7.65959      \\
    $S_1$                  & -7.64450 & -7.64340       & -7.73624 & -7.73554      & -7.75366 & -7.75321     & -7.74960 & -7.74934     & -7.73847 & -7.73834      \\
    $T_1$                  & -7.65893 & -7.65852       & -7.75180 & -7.75156      & -7.77343 & -7.77321     & -7.77729 & -7.77705     & -7.77845 & -7.77820      \\
    $S_0$                  & -7.78446 & -7.78434       & -7.87845 & -7.87833      & -7.87452 & -7.87438     & -7.84568 & -7.84552     & -7.81740 & -7.81720      \\
    \colrule
    \end{tabular}
    }
    \resizebox{.8\textwidth}{!}{%
    \begin{tabular}{l|cc|cc|cc|cc|c}
    \multirow{2}{*}{States}&\multicolumn{2}{c}{3.00}& \multicolumn{2}{c}{3.40}& \multicolumn{2}{c}{3.80}&\multicolumn{2}{c}{4.20}&\multirow{2}{*}{NPE /$mE_h$}  \\
                           & FCI & VQD                 & FCI & VQD                & FCI & VQD               & FCI & VQD                     &             \\ 
    \hline
    $T_3$                  & -7.68023 & -7.68008       & -7.68962 & -7.68948      & -7.69383 & -7.69369     & -7.69570 & -7.69556     & 0.06 \\
    $S_1$                  & -7.72461 & -7.72431       & -7.71235 & -7.71222      & -7.70444 & -7.70430     & -7.70032 & -7.70018     & 0.97 \\
    $T_1$                  & -7.77987 & -7.77961       & -7.78112 & -7.78086      & -7.78187 & -7.78160     & -7.78222 & -7.78195     & 0.18 \\
    $S_0$                  & -7.79884 & -7.79855       & -7.78950 & -7.78922      & -7.78535 & -7.78505     & -7.78359 & -7.78331     & 0.17 \\
    \colrule
    \end{tabular}
    }
    \caption{The values of the ADAPT-VQD energies at the computed bond lengths (in Hartree), and the NPE of recovered eigenstates (in milliHartree).}
    \label{tab:my_label}
    \end{table*}

\subsection{Spin-Restricted ADAPT-VQD} \label{section:adapt-vqd}
    With the singlet spin reference, the singlet ground state was first recovered as per the original ADAPT-VQE prescription.
    The spin references continue to be successful in isolating the relevant Hilbert space, and the next adaptively grown ansatz converged to the first singlet excited state $S_1$, avoiding the $T_1$ state.
    The energy was in good agreement with the exact solution, lying within 1.59 m$E_h$ across all bond lengths in both the $S_0$ and $S_1$ states (see Figure \ref{fig:ADAPT}).
    The $S_0$ and $S_1$ states appended, on average across calculated bond lengths, 10.9 and 14.1 excitation operators before convergence, culminating to ansatz circuit depths of 228.8 and 280.2 gates respectively.
    This was a 30 fold resource improvement over the UCCGSD and a 3 fold improvement over the 2-UpCCGSD circuit depth.
    For the $S_0$ ground state, the errors are similar to those from the 2-UpCCGSD-VQD calculation.
    For the $S_1$ excited state, the errors increase at shorter bond lengths, giving a NPE of 0.97 $mE_h$,  but still within chemical accuracy.
    Like the 2-UpCCGSD method, the $S^2$ value for all adaptive singlet states was 0.00.

    The triplet spin reference was also able to adaptively recover the triplet ground state $T_1$.
    Upon convergence however, the next excited eigenstate selected by the spin restricted ADAPT-VQD was not the next highest in energy (the doubly degenerate $T_2$ state), but the $T_3$ state above, a behaviour not seen in the two other tested ans\"atze which both recovered the $T_2$ state using the same reference circuit.
    We note that the states recovered with the ADAPT method share the same irreducible representation $A_1$, whereas the doubly degenerate $T_2$ state which the method missed had an irreducible representation of $E_1$.
    In the presence of molecular point group symmetry, due to the active selection of excitation operators in ADAPT, the set of eigenstates that a given starting reference can actually express is limited to those constructed from excitations that are symmetry-allowed from the reference.
    An excitation is symmetry-allowed if the direct product of the irreducible representations of the the orbitals involved in the excitation $i$, $j$, $a$, and $b$
    $(\Gamma_i \otimes \Gamma_j \otimes \Gamma_a \otimes \Gamma_b)$ contains the irreducible representation of the point group of the desired exited state.\cite{Lee2019, Kottmann2020a, Stanton1991ADP}.
    Consequently, addition of only symmetry-allowed single excitation operators (which can be thought of as rotations of the reference state in Hilbert space) in the presented demonstration rotates the state such that it remains in the totally symmetric manifold.
    Hence, the resulting ans\"atze were only able to recover $A_1$ states.
    
    In contrast, the ans\"{a}tze in the two preceding methods have sufficient flexibility to effectively mix reference occupied orbitals with virtual orbitals to obtain a different symmetry reference, which is then correlated.
    The greater degree of control over states of which symmetry is recovered in ADAPT-VQD is an advantageous feature over the other methods.
    Future work will investigate efficient methods for recovering any missing states.
    We suspect further contributions come from  the simultaneous presence of spin and point group symmetry.
    Moving forward, these symmetry observations must be extended to the irreducible representations of the direct product group $\Gamma \otimes SU(2) $, which simultaneously describes point group and spin symmetry.
    
    Apart from the convergence to the $T_3$ state, other performance metrics of the triplet calculations were comparable with the singlet calculations.
    Figure \ref{fig:ADAPT} shows that the triplet energy accuracy was very even across all computed geometries.
    The ansatz depth for calculating eigenstates in the triplet manifold even benefited from further reductions to the circuits for adaptive singlet states, circuits of only 192.2 and 213.4 gates deep on average for $T_1$ and $T_3$ before convergence. 
    The $S^2$ value was 2.00 for every triplet calculation.

\begin{figure}[h!]
        \includegraphics[scale=0.41]{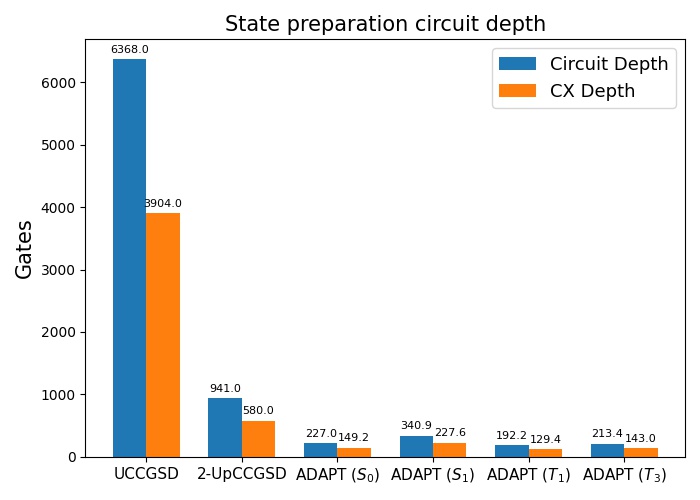}
        \caption{The state preparation circuit depth for the three different ansatz construction methods tested.
        Values reported for ADAPT-VQD are averages across the calculated bond lengths.}
        \label{fig:depth}
    \end{figure}

    One important consideration for ADAPT-VQD, which has not been studied here, is the circuit measurement overhead.
    The reductions in circuit depth for ADAPT come at the cost of more measurements at each ADAPT iteration for operator selection.
    For the system investigated here, each ADAPT iteration executes the expectation measurement of 30 excitation operator gradients, followed by a variational optimisation of the updated ansatz once a new operator has been selected and added.
    In contrast, the fixed-length ans\"atze only require expectation measurements for the variational optimisation of the parameters.
    As each expectation measurement is actually statistical in nature (see the Supplementary Material), the number of circuit executions and measurements in a real quantum device will be much higher for the ADAPT-VQD case than the fixed ans\"atze.
    Nonetheless, the significant reduction in circuit depth for encoding excited state wavefunctions with the ADAPT-VQD method makes it a very promising method for computing molecular excited state potential energy surfaces (PES) on NISQ devices, whereby the measurement overhead can be mitigated using, for example, parallel gradient computation.
    Table \ref{tb:resources} and Figure \ref{fig:depth} summarise the quantum circuit resources in the three methods investigated in this work.
    \begin{table}[h!]
    \centering
    \begin{tabular}{lccc}
        \toprule
        LiH (STO-3G) & Operators & Circuit depth & CX depth \\
        \colrule
        singlet UCCGSD (fixed)      & 135 & 6368 & 3904 \\
        singlet 2-UpCCGSD (fixed)   & 60 & 941 & 580 \\
        $S_0$ ADAPT (average)   & 10.9 & 227.0 & 149.2 \\
        $S_1$ ADAPT (average)   & 14.1 & 340.9 & 227.6 \\
        $T_1$ ADAPT (average)   & 7.3 & 192.2 & 129.4 \\
        $T_3$ ADAPT (average)   & 8.7 & 213.4 & 143.0 \\
        \colrule
    \end{tabular}\\
    \caption{Summary of quantum circuit resource needed to compute excited states of \textnormal{LiH} using spin restriction and different ans\"atze.
    The ansatz depth is fixed for both UCC-based techniques, whereas for the adaptively grown ansatz the values are averages across the 5 calculated bond lengths.
    The ansatz convergence threshold is taken to be $\epsilon=0.01$ for the ADAPT methods.
    }
    \label{tb:resources}
    \end{table}%
    \begin{figure}[h!]
        \centering
        \includegraphics[scale=0.41]{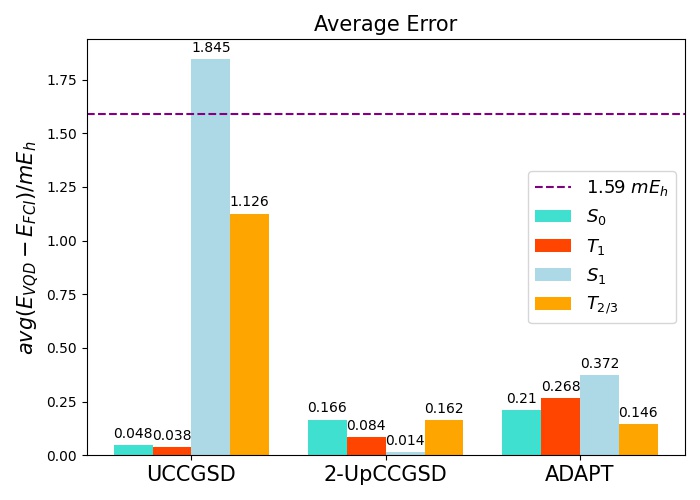}
        \caption{A comparison of the average error (in milliHartree), hence accuracy, between the three methods tested in this investigation.}
        \label{fig:accuracy}
    \end{figure}

\section{Conclusion}
\label{section:conclusion}
%%%% Key Summary %%%%%
We combined innovations from the variational quantum algorithm literature with quantum chemistry insight to demonstrate and systematically analyse an implementation of ADAPT-VQD for excited states.
Using quantum simulations on classical computers, we suggest that ADAPT-VQD will be more practical for calculating molecular excited states on near-term NISQ computers than other VQD implementations that use fixed-length ans\"atze such as UCCGSD and $k$-UpCCGSD, because it is able to prepare excited state wavefunctions on quantum computers with significantly reduced circuit depths than the other methods, while also maintaining chemical accuracy across different calculated nuclear geometries.
We also showed that segregated computation of states in the singlet and triplet manifold can be achieved with spin restrictions at the excitation operator and reference circuit level, which reduces the number of states that need to have converged before reaching higher states.

However, we also confirmed that in the ADAPT-VQD method, a reference circuit can only calculate eigenstates of a given irreducible representation of the molecular point group because the algorithm only selects excitations that are symmetry-allowed.
% These excitations are symmetry allowed if the direct product of the irreducible representations belonging to the orbitals $i$, $j$, $a$, and $b$ 
% $(\Gamma_i \otimes \Gamma_j \otimes \Gamma_a \otimes \Gamma_b)$ contains the irreducible representation of the point group of the desired exited state.
More work needs to be done on how one might recover missing states efficiently, as well as applying the suggested strategies to larger systems and more realistic basis sets.
% where the relationship of point group symmetry irreducible representation between spin references and the eigenstates they recover will be less clear.
% We suspect that due to the presence of spin and spatial symmetry, these symmetry observations must be extended to the irr. rep. of the direct product group $\Gamma \otimes SU(2) $ which simultaneously describes both.

As high quality excited state potential energy surfaces are necessary for the quantitative study of spectroscopic and dynamical properties of photochemical reactions, and suitable classical \textit{ab initio} methods still face severe scaling challenges, the largest excited state systems
% (at accurate coupled cluster approaches nearing chemical accuracy)
that could be investigated computationally with chemical accuracy are currently limited to small organic molecules.
It is therefore conceivable that one of the first demonstrations of quantum computers outperforming classical computers on a practically relevant task will be accurate calculations of excited states in molecules, as opposed to ground states of large complexes.  
In principle, with further development and rapidly improving hardware, a future more efficient iteration of the spin restricted ADAPT-VQD algorithm can perform molecular excited state calculations on near-term quantum computers without the need for quantum error correction.
Thus, this work lays the foundation for the possibility of simulating molecular excited states on real quantum computers in the much nearer future than anticipated.

\section*{Acknowledgements}
The authors would like to thank Cambridge Quantum Computing for their ongoing support.
H.H.S.C. would like to thank Daniel Marti-Dafcik, Nick Mayhall, Joonho Lee, Bill Huggins, Oscar Higgott and Yohei Ibe for helpful discussions.
J.S.-M. acknowledges the European Commission through the Marie Curie actions (\textit{AttoDNA}, FP8-MSCA-IF, grant n$^{\circ}$ 747662), and current support of a fellowship from La Caixa Foundation (ID 100010434) and from the European Union's Horizon 2020 research and innovation programme under the Marie Sk\l{}odowska-Curie grant agreement No 847648, fellowship code LCF/BQ/PI20/11760022.
Computation in this work was made possible by Imperial College London's Research Computing Service\cite{ICHPC}.
H.H.S.C. is funded by the Croucher Foundation, Hong Kong.

\bibliography{library.bib}

\providecommand*{\mcitethebibliography}{\thebibliography}
\csname @ifundefined\endcsname{endmcitethebibliography}
{\let\endmcitethebibliography\endthebibliography}{}
\begin{mcitethebibliography}{48}
\providecommand*{\natexlab}[1]{#1}
\providecommand*{\mciteSetBstSublistMode}[1]{}
\providecommand*{\mciteSetBstMaxWidthForm}[2]{}
\providecommand*{\mciteBstWouldAddEndPuncttrue}
  {\def\EndOfBibitem{\unskip.}}
\providecommand*{\mciteBstWouldAddEndPunctfalse}
  {\let\EndOfBibitem\relax}
\providecommand*{\mciteSetBstMidEndSepPunct}[3]{}
\providecommand*{\mciteSetBstSublistLabelBeginEnd}[3]{}
\providecommand*{\EndOfBibitem}{}
\mciteSetBstSublistMode{f}
\mciteSetBstMaxWidthForm{subitem}
{(\emph{\alph{mcitesubitemcount}})}
\mciteSetBstSublistLabelBeginEnd{\mcitemaxwidthsubitemform\space}
{\relax}{\relax}

\bibitem[Lindh and Gonz{\'{a}}lez(2020)]{Gonzalez2020}
R.~Lindh and L.~Gonz{\'{a}}lez, \emph{{Quantum Chemistry and Dynamics of
  Excited States: Methods and Applications}}, John Wiley and Sons, Ltd,
  2020\relax
\mciteBstWouldAddEndPuncttrue
\mciteSetBstMidEndSepPunct{\mcitedefaultmidpunct}
{\mcitedefaultendpunct}{\mcitedefaultseppunct}\relax
\EndOfBibitem
\bibitem[Eriksen \emph{et~al.}(2020)Eriksen, Anderson, Deustua, Ghanem, Hait,
  Hoffmann, Lee, Levine, Magoulas, Shen, Tubman, Whaley, Xu, Yao, Zhang, Alavi,
  Chan, Head-Gordon, Liu, Piecuch, Sharma, Ten-No, Umrigar, and
  Gauss]{Eriksen2020}
J.~J. Eriksen, T.~A. Anderson, J.~E. Deustua, K.~Ghanem, D.~Hait, M.~R.
  Hoffmann, S.~Lee, D.~S. Levine, I.~Magoulas, J.~Shen, N.~M. Tubman, K.~B.
  Whaley, E.~Xu, Y.~Yao, N.~Zhang, A.~Alavi, G.~K.~L. Chan, M.~Head-Gordon,
  W.~Liu, P.~Piecuch, S.~Sharma, S.~L. Ten-No, C.~J. Umrigar and J.~Gauss,
  \emph{Journal of Physical Chemistry Letters}, 2020, \textbf{11},
  8922--8929\relax
\mciteBstWouldAddEndPuncttrue
\mciteSetBstMidEndSepPunct{\mcitedefaultmidpunct}
{\mcitedefaultendpunct}{\mcitedefaultseppunct}\relax
\EndOfBibitem
\bibitem[McArdle \emph{et~al.}(2020)McArdle, Endo, Aspuru-Guzik, Benjamin, and
  Yuan]{McArdle2018}
S.~McArdle, S.~Endo, A.~Aspuru-Guzik, S.~C. Benjamin and X.~Yuan, \emph{Reviews
  of Modern Physics}, 2020, \textbf{92}, 015003\relax
\mciteBstWouldAddEndPuncttrue
\mciteSetBstMidEndSepPunct{\mcitedefaultmidpunct}
{\mcitedefaultendpunct}{\mcitedefaultseppunct}\relax
\EndOfBibitem
\bibitem[Cao \emph{et~al.}(2019)Cao, Romero, Olson, Degroote, Johnson,
  Kieferov{\'{a}}, Kivlichan, Menke, Peropadre, Sawaya, Sim, Veis, and
  Aspuru-Guzik]{Cao2019}
Y.~Cao, J.~Romero, J.~P. Olson, M.~Degroote, P.~D. Johnson, M.~Kieferov{\'{a}},
  I.~D. Kivlichan, T.~Menke, B.~Peropadre, N.~P.~D. Sawaya, S.~Sim, L.~Veis and
  A.~Aspuru-Guzik, \emph{Chemical Reviews}, 2019, \textbf{119},
  10856--10915\relax
\mciteBstWouldAddEndPuncttrue
\mciteSetBstMidEndSepPunct{\mcitedefaultmidpunct}
{\mcitedefaultendpunct}{\mcitedefaultseppunct}\relax
\EndOfBibitem
\bibitem[Bauer \emph{et~al.}(2020)Bauer, Bravyi, Motta, and Chan]{Bauer2020}
B.~Bauer, S.~Bravyi, M.~Motta and G.~K.~L. Chan, \emph{Chemical Reviews}, 2020,
  \textbf{120}, 12685--12717\relax
\mciteBstWouldAddEndPuncttrue
\mciteSetBstMidEndSepPunct{\mcitedefaultmidpunct}
{\mcitedefaultendpunct}{\mcitedefaultseppunct}\relax
\EndOfBibitem
\bibitem[Cerezo \emph{et~al.}(2020)Cerezo, Arrasmith, Babbush, Benjamin, Endo,
  Fujii, McClean, Mitarai, Yuan, Cincio, and Coles]{Cerezo2020a}
M.~Cerezo, A.~Arrasmith, R.~Babbush, S.~C. Benjamin, S.~Endo, K.~Fujii, J.~R.
  McClean, K.~Mitarai, X.~Yuan, L.~Cincio and P.~J. Coles, \emph{{Variational
  Quantum Algorithms}}, 2020, \url{http://arxiv.org/abs/2012.09265}\relax
\mciteBstWouldAddEndPuncttrue
\mciteSetBstMidEndSepPunct{\mcitedefaultmidpunct}
{\mcitedefaultendpunct}{\mcitedefaultseppunct}\relax
\EndOfBibitem
\bibitem[Lee \emph{et~al.}(2019)Lee, Huggins, Head-Gordon, and Whaley]{Lee2019}
J.~Lee, W.~J. Huggins, M.~Head-Gordon and K.~B. Whaley, \emph{Journal of
  Chemical Theory and Computation}, 2019, \textbf{15}, 311--324\relax
\mciteBstWouldAddEndPuncttrue
\mciteSetBstMidEndSepPunct{\mcitedefaultmidpunct}
{\mcitedefaultendpunct}{\mcitedefaultseppunct}\relax
\EndOfBibitem
\bibitem[Evangelista \emph{et~al.}(2019)Evangelista, Chan, and
  Scuseria]{Evangelista2019}
F.~A. Evangelista, G.~K.~L. Chan and G.~E. Scuseria, \emph{The Journal of
  Chemical Physics}, 2019, \textbf{151}, 244112\relax
\mciteBstWouldAddEndPuncttrue
\mciteSetBstMidEndSepPunct{\mcitedefaultmidpunct}
{\mcitedefaultendpunct}{\mcitedefaultseppunct}\relax
\EndOfBibitem
\bibitem[Bartlett and Musia{\l}(2007)]{Bartlett2007}
R.~J. Bartlett and M.~Musia{\l}, \emph{Reviews of Modern Physics}, 2007,
  \textbf{79}, 291--352\relax
\mciteBstWouldAddEndPuncttrue
\mciteSetBstMidEndSepPunct{\mcitedefaultmidpunct}
{\mcitedefaultendpunct}{\mcitedefaultseppunct}\relax
\EndOfBibitem
\bibitem[Higgott \emph{et~al.}(2019)Higgott, Wang, and Brierley]{Higgott2019}
O.~Higgott, D.~Wang and S.~Brierley, \emph{Quantum}, 2019, \textbf{3},
  156\relax
\mciteBstWouldAddEndPuncttrue
\mciteSetBstMidEndSepPunct{\mcitedefaultmidpunct}
{\mcitedefaultendpunct}{\mcitedefaultseppunct}\relax
\EndOfBibitem
\bibitem[Peruzzo \emph{et~al.}(2014)Peruzzo, McClean, Shadbolt, Yung, Zhou,
  Love, Aspuru-Guzik, and O'Brien]{Peruzzo2014}
A.~Peruzzo, J.~McClean, P.~Shadbolt, M.-H. Yung, X.-Q. Zhou, P.~J. Love,
  A.~Aspuru-Guzik and J.~L. O'Brien, \emph{Nature Communications}, 2014,
  \textbf{5}, 4213\relax
\mciteBstWouldAddEndPuncttrue
\mciteSetBstMidEndSepPunct{\mcitedefaultmidpunct}
{\mcitedefaultendpunct}{\mcitedefaultseppunct}\relax
\EndOfBibitem
\bibitem[Kandala \emph{et~al.}(2017)Kandala, Mezzacapo, Temme, Takita, Brink,
  Chow, and Gambetta]{Kandala2017}
A.~Kandala, A.~Mezzacapo, K.~Temme, M.~Takita, M.~Brink, J.~M. Chow and J.~M.
  Gambetta, \emph{Nature}, 2017, \textbf{549}, 242--246\relax
\mciteBstWouldAddEndPuncttrue
\mciteSetBstMidEndSepPunct{\mcitedefaultmidpunct}
{\mcitedefaultendpunct}{\mcitedefaultseppunct}\relax
\EndOfBibitem
\bibitem[Colless \emph{et~al.}(2018)Colless, Ramasesh, Dahlen, Blok,
  Kimchi-Schwartz, McClean, Carter, de~Jong, and Siddiqi]{Colless2018a}
J.~I. Colless, V.~V. Ramasesh, D.~Dahlen, M.~S. Blok, M.~E. Kimchi-Schwartz,
  J.~R. McClean, J.~Carter, W.~A. de~Jong and I.~Siddiqi, \emph{Physical Review
  X}, 2018, \textbf{8}, 011021\relax
\mciteBstWouldAddEndPuncttrue
\mciteSetBstMidEndSepPunct{\mcitedefaultmidpunct}
{\mcitedefaultendpunct}{\mcitedefaultseppunct}\relax
\EndOfBibitem
\bibitem[Hempel \emph{et~al.}(2018)Hempel, Maier, Romero, McClean, Monz, Shen,
  Jurcevic, Lanyon, Love, Babbush, Aspuru-Guzik, Blatt, and Roos]{Hempel2018}
C.~Hempel, C.~Maier, J.~Romero, J.~McClean, T.~Monz, H.~Shen, P.~Jurcevic,
  B.~P. Lanyon, P.~Love, R.~Babbush, A.~Aspuru-Guzik, R.~Blatt and C.~F. Roos,
  \emph{Physical Review X}, 2018, \textbf{8}, 031022\relax
\mciteBstWouldAddEndPuncttrue
\mciteSetBstMidEndSepPunct{\mcitedefaultmidpunct}
{\mcitedefaultendpunct}{\mcitedefaultseppunct}\relax
\EndOfBibitem
\bibitem[O'Malley \emph{et~al.}(2016)O'Malley, Babbush, Kivlichan, Romero,
  McClean, Barends, Kelly, Roushan, Tranter, Ding, Campbell, Chen, Chen,
  Chiaro, Dunsworth, Fowler, Jeffrey, Lucero, Megrant, Mutus, Neeley, Neill,
  Quintana, Sank, Vainsencher, Wenner, White, Coveney, Love, Neven,
  Aspuru-Guzik, and Martinis]{OMalley2016}
P.~J.~J. O'Malley, R.~Babbush, I.~D. Kivlichan, J.~Romero, J.~R. McClean,
  R.~Barends, J.~Kelly, P.~Roushan, A.~Tranter, N.~Ding, B.~Campbell, Y.~Chen,
  Z.~Chen, B.~Chiaro, A.~Dunsworth, A.~G. Fowler, E.~Jeffrey, E.~Lucero,
  A.~Megrant, J.~Y. Mutus, M.~Neeley, C.~Neill, C.~Quintana, D.~Sank,
  A.~Vainsencher, J.~Wenner, T.~C. White, P.~V. Coveney, P.~J. Love, H.~Neven,
  A.~Aspuru-Guzik and J.~M. Martinis, \emph{Physical Review X}, 2016,
  \textbf{6}, 031007\relax
\mciteBstWouldAddEndPuncttrue
\mciteSetBstMidEndSepPunct{\mcitedefaultmidpunct}
{\mcitedefaultendpunct}{\mcitedefaultseppunct}\relax
\EndOfBibitem
\bibitem[Kokail \emph{et~al.}(2019)Kokail, Maier, van Bijnen, Brydges, Joshi,
  Jurcevic, Muschik, Silvi, Blatt, Roos, and Zoller]{Kokail2019}
C.~Kokail, C.~Maier, R.~van Bijnen, T.~Brydges, M.~K. Joshi, P.~Jurcevic, C.~A.
  Muschik, P.~Silvi, R.~Blatt, C.~F. Roos and P.~Zoller, \emph{Nature}, 2019,
  \textbf{569}, 355--360\relax
\mciteBstWouldAddEndPuncttrue
\mciteSetBstMidEndSepPunct{\mcitedefaultmidpunct}
{\mcitedefaultendpunct}{\mcitedefaultseppunct}\relax
\EndOfBibitem
\bibitem[Ganzhorn \emph{et~al.}(2019)Ganzhorn, Egger, Barkoutsos, Ollitrault,
  Salis, Moll, Roth, Fuhrer, Mueller, Woerner, Tavernelli, and
  Filipp]{Ganzhorn2019}
M.~Ganzhorn, D.~Egger, P.~Barkoutsos, P.~Ollitrault, G.~Salis, N.~Moll,
  M.~Roth, A.~Fuhrer, P.~Mueller, S.~Woerner, I.~Tavernelli and S.~Filipp,
  \emph{Physical Review Applied}, 2019, \textbf{11}, 044092\relax
\mciteBstWouldAddEndPuncttrue
\mciteSetBstMidEndSepPunct{\mcitedefaultmidpunct}
{\mcitedefaultendpunct}{\mcitedefaultseppunct}\relax
\EndOfBibitem
\bibitem[{AI Quantum}(2020)]{AIQuantum2020}
G.~{AI Quantum}, \emph{Science}, 2020, \textbf{369}, 1084--1089\relax
\mciteBstWouldAddEndPuncttrue
\mciteSetBstMidEndSepPunct{\mcitedefaultmidpunct}
{\mcitedefaultendpunct}{\mcitedefaultseppunct}\relax
\EndOfBibitem
\bibitem[Ibe \emph{et~al.}(2020)Ibe, Nakagawa, Yamamoto, Mitarai, Gao, and
  Kobayashi]{Ibe2020}
Y.~Ibe, Y.~O. Nakagawa, T.~Yamamoto, K.~Mitarai, Q.~Gao and T.~Kobayashi,
  \emph{{Calculating transition amplitudes by variational quantum
  eigensolvers}}, 2020, \url{http://arxiv.org/abs/2002.11724}\relax
\mciteBstWouldAddEndPuncttrue
\mciteSetBstMidEndSepPunct{\mcitedefaultmidpunct}
{\mcitedefaultendpunct}{\mcitedefaultseppunct}\relax
\EndOfBibitem
\bibitem[McArdle \emph{et~al.}(2018)McArdle, Jones, Endo, Li, Benjamin, and
  Yuan]{McArdle2018a}
S.~McArdle, T.~Jones, S.~Endo, Y.~Li, S.~Benjamin and X.~Yuan, \emph{npj
  Quantum Information}, 2018, \textbf{5}, 1--13\relax
\mciteBstWouldAddEndPuncttrue
\mciteSetBstMidEndSepPunct{\mcitedefaultmidpunct}
{\mcitedefaultendpunct}{\mcitedefaultseppunct}\relax
\EndOfBibitem
\bibitem[Endo \emph{et~al.}(2018)Endo, Benjamin, and Li]{Endo2018}
S.~Endo, S.~C. Benjamin and Y.~Li, \emph{Physical Review X}, 2018, \textbf{8},
  31027\relax
\mciteBstWouldAddEndPuncttrue
\mciteSetBstMidEndSepPunct{\mcitedefaultmidpunct}
{\mcitedefaultendpunct}{\mcitedefaultseppunct}\relax
\EndOfBibitem
\bibitem[McClean \emph{et~al.}(2017)McClean, Kimchi-Schwartz, Carter, and
  de~Jong]{McClean2017}
J.~R. McClean, M.~E. Kimchi-Schwartz, J.~Carter and W.~A. de~Jong, \emph{Phys.
  Rev. A}, 2017, \textbf{95}, 042308\relax
\mciteBstWouldAddEndPuncttrue
\mciteSetBstMidEndSepPunct{\mcitedefaultmidpunct}
{\mcitedefaultendpunct}{\mcitedefaultseppunct}\relax
\EndOfBibitem
\bibitem[Santagati \emph{et~al.}(2018)Santagati, Wang, Gentile, Paesani, Wiebe,
  McClean, Morley-Short, Shadbolt, Bonneau, Silverstone, Tew, Zhou,
  O{\textquoteright}Brien, and Thompson]{Santagati2018}
R.~Santagati, J.~Wang, A.~A. Gentile, S.~Paesani, N.~Wiebe, J.~R. McClean,
  S.~Morley-Short, P.~J. Shadbolt, D.~Bonneau, J.~W. Silverstone, D.~P. Tew,
  X.~Zhou, J.~L. O{\textquoteright}Brien and M.~G. Thompson, \emph{Science
  Advances}, 2018, \textbf{4}, eaap9646\relax
\mciteBstWouldAddEndPuncttrue
\mciteSetBstMidEndSepPunct{\mcitedefaultmidpunct}
{\mcitedefaultendpunct}{\mcitedefaultseppunct}\relax
\EndOfBibitem
\bibitem[Bauman \emph{et~al.}(2019)Bauman, Low, and Kowalski]{Bauman2019}
N.~P. Bauman, G.~H. Low and K.~Kowalski, \emph{The Journal of Chemical
  Physics}, 2019, \textbf{151}, 234114\relax
\mciteBstWouldAddEndPuncttrue
\mciteSetBstMidEndSepPunct{\mcitedefaultmidpunct}
{\mcitedefaultendpunct}{\mcitedefaultseppunct}\relax
\EndOfBibitem
\bibitem[Bauman \emph{et~al.}(2021)Bauman, Liu, Bylaska, Krishnamoorthy, Low,
  Granade, Wiebe, Baker, Peng, Roetteler, Troyer, and Kowalski]{Bauman2021}
N.~P. Bauman, H.~Liu, E.~J. Bylaska, S.~Krishnamoorthy, G.~H. Low, C.~E.
  Granade, N.~Wiebe, N.~A. Baker, B.~Peng, M.~Roetteler, M.~Troyer and
  K.~Kowalski, \emph{Journal of Chemical Theory and Computation}, 2021,
  \textbf{17}, 201--210\relax
\mciteBstWouldAddEndPuncttrue
\mciteSetBstMidEndSepPunct{\mcitedefaultmidpunct}
{\mcitedefaultendpunct}{\mcitedefaultseppunct}\relax
\EndOfBibitem
\bibitem[Sugisaki \emph{et~al.}(2021)Sugisaki, Toyota, Sato, Shiomi, and
  Takui]{Sugisaki2021}
K.~Sugisaki, K.~Toyota, K.~Sato, D.~Shiomi and T.~Takui, \emph{The Journal of
  Physical Chemistry Letters}, 2021, \textbf{12}, 2880--2885\relax
\mciteBstWouldAddEndPuncttrue
\mciteSetBstMidEndSepPunct{\mcitedefaultmidpunct}
{\mcitedefaultendpunct}{\mcitedefaultseppunct}\relax
\EndOfBibitem
\bibitem[Greene-Diniz and Ramo(2020)]{Greene-Diniz2019}
G.~Greene-Diniz and D.~M. Ramo, \emph{{Generalized unitary coupled cluster
  excitations for multireference molecular states optimized by the Variational
  Quantum Eigensolver}}, 2020, \url{http://arxiv.org/abs/1910.05168}\relax
\mciteBstWouldAddEndPuncttrue
\mciteSetBstMidEndSepPunct{\mcitedefaultmidpunct}
{\mcitedefaultendpunct}{\mcitedefaultseppunct}\relax
\EndOfBibitem
\bibitem[Pancuz(1979)]{Pancuz1979}
R.~Pancuz, \emph{Spin Eigenfunctions: Construction and Use}, Springer US, 1st
  edn., 1979\relax
\mciteBstWouldAddEndPuncttrue
\mciteSetBstMidEndSepPunct{\mcitedefaultmidpunct}
{\mcitedefaultendpunct}{\mcitedefaultseppunct}\relax
\EndOfBibitem
\bibitem[Grimsley \emph{et~al.}(2019)Grimsley, Economou, Barnes, and
  Mayhall]{Grimsley2019}
H.~R. Grimsley, S.~E. Economou, E.~Barnes and N.~J. Mayhall, \emph{Nature
  Communications}, 2019, \textbf{10}, 1--11\relax
\mciteBstWouldAddEndPuncttrue
\mciteSetBstMidEndSepPunct{\mcitedefaultmidpunct}
{\mcitedefaultendpunct}{\mcitedefaultseppunct}\relax
\EndOfBibitem
\bibitem[Tang \emph{et~al.}(2019)Tang, Barnes, Grimsley, Mayhall, and
  Economou]{Tang2019}
H.~L. Tang, E.~Barnes, H.~R. Grimsley, N.~J. Mayhall and S.~E. Economou,
  \emph{{qubit-ADAPT-VQE: An adaptive algorithm for constructing
  hardware-efficient ansatze on a quantum processor}}, 2019,
  \url{http://arxiv.org/abs/1911.10205}\relax
\mciteBstWouldAddEndPuncttrue
\mciteSetBstMidEndSepPunct{\mcitedefaultmidpunct}
{\mcitedefaultendpunct}{\mcitedefaultseppunct}\relax
\EndOfBibitem
\bibitem[Rattew \emph{et~al.}(2019)Rattew, Hu, Pistoia, Chen, and
  Wood]{Rattew2019}
A.~G. Rattew, S.~Hu, M.~Pistoia, R.~Chen and S.~Wood, \emph{{A Domain-agnostic,
  Noise-resistant, Hardware-efficient Evolutionary Variational Quantum
  Eigensolver}}, 2019, \url{http://arxiv.org/abs/1910.09694}\relax
\mciteBstWouldAddEndPuncttrue
\mciteSetBstMidEndSepPunct{\mcitedefaultmidpunct}
{\mcitedefaultendpunct}{\mcitedefaultseppunct}\relax
\EndOfBibitem
\bibitem[Claudino \emph{et~al.}(2020)Claudino, Wright, McCaskey, and
  Humble]{Claudino2020}
D.~Claudino, J.~Wright, A.~J. McCaskey and T.~S. Humble, \emph{Frontiers in
  Chemistry}, 2020, \textbf{8}, 1--12\relax
\mciteBstWouldAddEndPuncttrue
\mciteSetBstMidEndSepPunct{\mcitedefaultmidpunct}
{\mcitedefaultendpunct}{\mcitedefaultseppunct}\relax
\EndOfBibitem
\bibitem[Kottmann \emph{et~al.}(2021)Kottmann, Anand, and
  Aspuru-Guzik]{Kottmann2020a}
J.~S. Kottmann, A.~Anand and A.~Aspuru-Guzik, \emph{Chem. Sci.}, 2021,  Advance
  article\relax
\mciteBstWouldAddEndPuncttrue
\mciteSetBstMidEndSepPunct{\mcitedefaultmidpunct}
{\mcitedefaultendpunct}{\mcitedefaultseppunct}\relax
\EndOfBibitem
\bibitem[Chan(2021)]{chan_2021}
H.~H.~S. Chan, \emph{\url{https://github.com/hanschanhs/QEBAB}}, 2021,
  \url{https://github.com/hanschanhs/QEBAB}\relax
\mciteBstWouldAddEndPuncttrue
\mciteSetBstMidEndSepPunct{\mcitedefaultmidpunct}
{\mcitedefaultendpunct}{\mcitedefaultseppunct}\relax
\EndOfBibitem
\bibitem[Romero \emph{et~al.}(2017)Romero, Babbush, McClean, Hempel, Love, and
  Aspuru-Guzik]{Romero2019}
J.~Romero, R.~Babbush, J.~R. McClean, C.~Hempel, P.~Love and A.~Aspuru-Guzik,
  \emph{Quantum Science and Technology}, 2017, \textbf{4}, 1--22\relax
\mciteBstWouldAddEndPuncttrue
\mciteSetBstMidEndSepPunct{\mcitedefaultmidpunct}
{\mcitedefaultendpunct}{\mcitedefaultseppunct}\relax
\EndOfBibitem
\bibitem[McClean \emph{et~al.}(2020)McClean, Rubin, Sung, Kivlichan,
  Bonet-Monroig, Cao, Dai, Fried, Gidney, Gimby, Gokhale, H{\"{a}}ner,
  Hardikar, Havl{\'{i}}{\v{c}}ek, Higgott, Huang, Izaac, Jiang, Liu, McArdle,
  Neeley, O'Brien, O'Gorman, Ozfidan, Radin, Romero, Sawaya, Senjean, Setia,
  Sim, Steiger, Steudtner, Sun, Sun, Wang, Zhang, and Babbush]{McClean2017a}
J.~R. McClean, N.~C. Rubin, K.~J. Sung, I.~D. Kivlichan, X.~Bonet-Monroig,
  Y.~Cao, C.~Dai, E.~S. Fried, C.~Gidney, B.~Gimby, P.~Gokhale, T.~H{\"{a}}ner,
  T.~Hardikar, V.~Havl{\'{i}}{\v{c}}ek, O.~Higgott, C.~Huang, J.~Izaac,
  Z.~Jiang, X.~Liu, S.~McArdle, M.~Neeley, T.~O'Brien, B.~O'Gorman, I.~Ozfidan,
  M.~D. Radin, J.~Romero, N.~P.~D. Sawaya, B.~Senjean, K.~Setia, S.~Sim, D.~S.
  Steiger, M.~Steudtner, Q.~Sun, W.~Sun, D.~Wang, F.~Zhang and R.~Babbush,
  \emph{Quantum Science and Technology}, 2020, \textbf{5}, 034014\relax
\mciteBstWouldAddEndPuncttrue
\mciteSetBstMidEndSepPunct{\mcitedefaultmidpunct}
{\mcitedefaultendpunct}{\mcitedefaultseppunct}\relax
\EndOfBibitem
\bibitem[Sun \emph{et~al.}(2018)Sun, Berkelbach, Blunt, Booth, Guo, Li, Liu,
  McClain, Sayfutyarova, Sharma, Wouters, and Chan]{Sun2018}
Q.~Sun, T.~C. Berkelbach, N.~S. Blunt, G.~H. Booth, S.~Guo, Z.~Li, J.~Liu,
  J.~D. McClain, E.~R. Sayfutyarova, S.~Sharma, S.~Wouters and G.~K.~L. Chan,
  \emph{WIREs Computational Molecular Science}, 2018, \textbf{8}, 1--15\relax
\mciteBstWouldAddEndPuncttrue
\mciteSetBstMidEndSepPunct{\mcitedefaultmidpunct}
{\mcitedefaultendpunct}{\mcitedefaultseppunct}\relax
\EndOfBibitem
\bibitem[Sun(2015)]{Libcint}
Q.~Sun, \emph{Journal of Computational Chemistry}, 2015, \textbf{36},
  1664--1671\relax
\mciteBstWouldAddEndPuncttrue
\mciteSetBstMidEndSepPunct{\mcitedefaultmidpunct}
{\mcitedefaultendpunct}{\mcitedefaultseppunct}\relax
\EndOfBibitem
\bibitem[McClean \emph{et~al.}(2020)McClean, Faulstich, Zhu, O'Gorman, Qiu,
  White, Babbush, and Lin]{McClean2020}
J.~R. McClean, F.~M. Faulstich, Q.~Zhu, B.~O'Gorman, Y.~Qiu, S.~R. White,
  R.~Babbush and L.~Lin, \emph{New Journal of Physics}, 2020, \textbf{22},
  093015\relax
\mciteBstWouldAddEndPuncttrue
\mciteSetBstMidEndSepPunct{\mcitedefaultmidpunct}
{\mcitedefaultendpunct}{\mcitedefaultseppunct}\relax
\EndOfBibitem
\bibitem[Sokolov \emph{et~al.}(2020)Sokolov, Barkoutsos, Ollitrault, Greenberg,
  Rice, Pistoia, and Tavernelli]{Sokolov2020}
I.~O. Sokolov, P.~K. Barkoutsos, P.~J. Ollitrault, D.~Greenberg, J.~Rice,
  M.~Pistoia and I.~Tavernelli, \emph{The Journal of Chemical Physics}, 2020,
  \textbf{152}, 124107\relax
\mciteBstWouldAddEndPuncttrue
\mciteSetBstMidEndSepPunct{\mcitedefaultmidpunct}
{\mcitedefaultendpunct}{\mcitedefaultseppunct}\relax
\EndOfBibitem
\bibitem[Suzuki \emph{et~al.}(2020)Suzuki, Kawase, Masumura, Hiraga, Nakadai,
  Chen, Nakanishi, Mitarai, Imai, Tamiya, Yamamoto, Yan, Kawakubo, Nakagawa,
  Ibe, Zhang, Yamashita, Yoshimura, Hayashi, and Fujii]{suzuki2020qulacs}
Y.~Suzuki, Y.~Kawase, Y.~Masumura, Y.~Hiraga, M.~Nakadai, J.~Chen, K.~M.
  Nakanishi, K.~Mitarai, R.~Imai, S.~Tamiya, T.~Yamamoto, T.~Yan, T.~Kawakubo,
  Y.~O. Nakagawa, Y.~Ibe, Y.~Zhang, H.~Yamashita, H.~Yoshimura, A.~Hayashi and
  K.~Fujii, \emph{Qulacs: a fast and versatile quantum circuit simulator for
  research purpose}, 2020, \url{https://arxiv.org/abs/2011.13524}\relax
\mciteBstWouldAddEndPuncttrue
\mciteSetBstMidEndSepPunct{\mcitedefaultmidpunct}
{\mcitedefaultendpunct}{\mcitedefaultseppunct}\relax
\EndOfBibitem
\bibitem[Virtanen \emph{et~al.}(2020)Virtanen, Gommers, Oliphant, Haberland,
  Reddy, Cournapeau, Burovski, Peterson, Weckesser, Bright, van~der Walt,
  Brett, Wilson, Millman, Mayorov, Nelson, Jones, Kern, Larson, Carey, Polat,
  Feng, Moore, VanderPlas, Laxalde, Perktold, Cimrman, Henriksen, Quintero,
  Harris, Archibald, Ribeiro, Pedregosa, van Mulbregt, Vijaykumar, Bardelli,
  Rothberg, Hilboll, Kloeckner, Scopatz, Lee, Rokem, Woods, Fulton, Masson,
  H{\"{a}}ggstr{\"{o}}m, Fitzgerald, Nicholson, Hagen, Pasechnik, Olivetti,
  Martin, Wieser, Silva, Lenders, Wilhelm, Young, Price, Ingold, Allen, Lee,
  Audren, Probst, Dietrich, Silterra, Webber, Slavi{\v{c}}, Nothman, Buchner,
  Kulick, Sch{\"{o}}nberger, {de Miranda Cardoso}, Reimer, Harrington,
  Rodr{\'{i}}guez, Nunez-Iglesias, Kuczynski, Tritz, Thoma, Newville,
  K{\"{u}}mmerer, Bolingbroke, Tartre, Pak, Smith, Nowaczyk, Shebanov, Pavlyk,
  Brodtkorb, Lee, McGibbon, Feldbauer, Lewis, Tygier, Sievert, Vigna, Peterson,
  More, Pudlik, Oshima, Pingel, Robitaille, Spura, Jones, Cera, Leslie, Zito,
  Krauss, Upadhyay, Halchenko, and V{\'{a}}zquez-Baeza]{Virtanen2020}
P.~Virtanen, R.~Gommers, T.~E. Oliphant, M.~Haberland, T.~Reddy, D.~Cournapeau,
  E.~Burovski, P.~Peterson, W.~Weckesser, J.~Bright, S.~J. van~der Walt,
  M.~Brett, J.~Wilson, K.~J. Millman, N.~Mayorov, A.~R. Nelson, E.~Jones,
  R.~Kern, E.~Larson, C.~J. Carey, Ä.~Polat, Y.~Feng, E.~W. Moore,
  J.~VanderPlas, D.~Laxalde, J.~Perktold, R.~Cimrman, I.~Henriksen, E.~A.
  Quintero, C.~R. Harris, A.~M. Archibald, A.~H. Ribeiro, F.~Pedregosa, P.~van
  Mulbregt, A.~Vijaykumar, A.~P. Bardelli, A.~Rothberg, A.~Hilboll,
  A.~Kloeckner, A.~Scopatz, A.~Lee, A.~Rokem, C.~N. Woods, C.~Fulton,
  C.~Masson, C.~H{\"{a}}ggstr{\"{o}}m, C.~Fitzgerald, D.~A. Nicholson, D.~R.
  Hagen, D.~V. Pasechnik, E.~Olivetti, E.~Martin, E.~Wieser, F.~Silva,
  F.~Lenders, F.~Wilhelm, G.~Young, G.~A. Price, G.~L. Ingold, G.~E. Allen,
  G.~R. Lee, H.~Audren, I.~Probst, J.~P. Dietrich, J.~Silterra, J.~T. Webber,
  J.~Slavi{\v{c}}, J.~Nothman, J.~Buchner, J.~Kulick, J.~L. Sch{\"{o}}nberger,
  J.~V. {de Miranda Cardoso}, J.~Reimer, J.~Harrington, J.~L.~C.
  Rodr{\'{i}}guez, J.~Nunez-Iglesias, J.~Kuczynski, K.~Tritz, M.~Thoma,
  M.~Newville, M.~K{\"{u}}mmerer, M.~Bolingbroke, M.~Tartre, M.~Pak, N.~J.
  Smith, N.~Nowaczyk, N.~Shebanov, O.~Pavlyk, P.~A. Brodtkorb, P.~Lee, R.~T.
  McGibbon, R.~Feldbauer, S.~Lewis, S.~Tygier, S.~Sievert, S.~Vigna,
  S.~Peterson, S.~More, T.~Pudlik, T.~Oshima, T.~J. Pingel, T.~P. Robitaille,
  T.~Spura, T.~R. Jones, T.~Cera, T.~Leslie, T.~Zito, T.~Krauss, U.~Upadhyay,
  Y.~O. Halchenko and Y.~V{\'{a}}zquez-Baeza, \emph{Nature Methods}, 2020,
  \textbf{17}, 261--272\relax
\mciteBstWouldAddEndPuncttrue
\mciteSetBstMidEndSepPunct{\mcitedefaultmidpunct}
{\mcitedefaultendpunct}{\mcitedefaultseppunct}\relax
\EndOfBibitem
\bibitem[Arute \emph{et~al.}(2019)Arute, Arya, Babbush, Bacon, Bardin, Barends,
  Biswas, Boixo, Brandao, Buell, Burkett, Chen, Chen, Chiaro, Collins,
  Courtney, Dunsworth, Farhi, Foxen, Fowler, Gidney, Giustina, Graff, Guerin,
  Habegger, Harrigan, Hartmann, Ho, Hoffmann, Huang, Humble, Isakov, Jeffrey,
  Jiang, Kafri, Kechedzhi, Kelly, Klimov, Knysh, Korotkov, Kostritsa, Landhuis,
  Lindmark, Lucero, Lyakh, Mandr{\`{a}}, McClean, McEwen, Megrant, Mi,
  Michielsen, Mohseni, Mutus, Naaman, Neeley, Neill, Niu, Ostby, Petukhov,
  Platt, Quintana, Rieffel, Roushan, Rubin, Sank, Satzinger, Smelyanskiy, Sung,
  Trevithick, Vainsencher, Villalonga, White, Yao, Yeh, Zalcman, Neven, and
  Martinis]{Arute2019}
F.~Arute, K.~Arya, R.~Babbush, D.~Bacon, J.~C. Bardin, R.~Barends, R.~Biswas,
  S.~Boixo, F.~G. Brandao, D.~A. Buell, B.~Burkett, Y.~Chen, Z.~Chen,
  B.~Chiaro, R.~Collins, W.~Courtney, A.~Dunsworth, E.~Farhi, B.~Foxen,
  A.~Fowler, C.~Gidney, M.~Giustina, R.~Graff, K.~Guerin, S.~Habegger, M.~P.
  Harrigan, M.~J. Hartmann, A.~Ho, M.~Hoffmann, T.~Huang, T.~S. Humble, S.~V.
  Isakov, E.~Jeffrey, Z.~Jiang, D.~Kafri, K.~Kechedzhi, J.~Kelly, P.~V. Klimov,
  S.~Knysh, A.~Korotkov, F.~Kostritsa, D.~Landhuis, M.~Lindmark, E.~Lucero,
  D.~Lyakh, S.~Mandr{\`{a}}, J.~R. McClean, M.~McEwen, A.~Megrant, X.~Mi,
  K.~Michielsen, M.~Mohseni, J.~Mutus, O.~Naaman, M.~Neeley, C.~Neill, M.~Y.
  Niu, E.~Ostby, A.~Petukhov, J.~C. Platt, C.~Quintana, E.~G. Rieffel,
  P.~Roushan, N.~C. Rubin, D.~Sank, K.~J. Satzinger, V.~Smelyanskiy, K.~J.
  Sung, M.~D. Trevithick, A.~Vainsencher, B.~Villalonga, T.~White, Z.~J. Yao,
  P.~Yeh, A.~Zalcman, H.~Neven and J.~M. Martinis, \emph{Nature}, 2019,
  \textbf{574}, 505--510\relax
\mciteBstWouldAddEndPuncttrue
\mciteSetBstMidEndSepPunct{\mcitedefaultmidpunct}
{\mcitedefaultendpunct}{\mcitedefaultseppunct}\relax
\EndOfBibitem
\bibitem[Docken and Hinze(1972)]{Docken1972}
K.~Docken and J.~Hinze, \emph{Journal of Chemical Physics}, 1972, \textbf{57},
  4928\relax
\mciteBstWouldAddEndPuncttrue
\mciteSetBstMidEndSepPunct{\mcitedefaultmidpunct}
{\mcitedefaultendpunct}{\mcitedefaultseppunct}\relax
\EndOfBibitem
\bibitem[Cowtan \emph{et~al.}(2020)Cowtan, Dilkes, Duncan, Simmons, and
  Sivarajah]{Cowtan2019}
A.~Cowtan, S.~Dilkes, R.~Duncan, W.~Simmons and S.~Sivarajah, \emph{Electronic
  Proceedings in Theoretical Computer Science}, 2020, \textbf{318},
  214--229\relax
\mciteBstWouldAddEndPuncttrue
\mciteSetBstMidEndSepPunct{\mcitedefaultmidpunct}
{\mcitedefaultendpunct}{\mcitedefaultseppunct}\relax
\EndOfBibitem
\bibitem[Cowtan \emph{et~al.}(2020)Cowtan, Simmons, and
  Duncan]{cowtan2020generic}
A.~Cowtan, W.~Simmons and R.~Duncan, \emph{A Generic Compilation Strategy for
  the Unitary Coupled Cluster Ansatz}, 2020,
  \url{https://arxiv.org/abs/2007.10515}\relax
\mciteBstWouldAddEndPuncttrue
\mciteSetBstMidEndSepPunct{\mcitedefaultmidpunct}
{\mcitedefaultendpunct}{\mcitedefaultseppunct}\relax
\EndOfBibitem
\bibitem[Stanton \emph{et~al.}(1991)Stanton, Gauss, Watts, and
  Bartlett]{Stanton1991ADP}
J.~Stanton, J.~Gauss, J.~Watts and R.~Bartlett, \emph{Journal of Chemical
  Physics}, 1991, \textbf{94}, 4334--4345\relax
\mciteBstWouldAddEndPuncttrue
\mciteSetBstMidEndSepPunct{\mcitedefaultmidpunct}
{\mcitedefaultendpunct}{\mcitedefaultseppunct}\relax
\EndOfBibitem
\bibitem[ICH()]{ICHPC}
\emph{Imperial College Research Computing Service,
  DOI:\href{https://www.imperial.ac.uk/admin-services/ict/self-service/research-support/rcs/}{10.14469/hpc/2232}}\relax
\mciteBstWouldAddEndPuncttrue
\mciteSetBstMidEndSepPunct{\mcitedefaultmidpunct}
{\mcitedefaultendpunct}{\mcitedefaultseppunct}\relax
\EndOfBibitem
\end{mcitethebibliography}


\providecommand*{\mcitethebibliography}{\thebibliography}
\csname @ifundefined\endcsname{endmcitethebibliography}
{\let\endmcitethebibliography\endthebibliography}{}
\begin{mcitethebibliography}{8}
\providecommand*{\natexlab}[1]{#1}
\providecommand*{\mciteSetBstSublistMode}[1]{}
\providecommand*{\mciteSetBstMaxWidthForm}[2]{}
\providecommand*{\mciteBstWouldAddEndPuncttrue}
  {\def\EndOfBibitem{\unskip.}}
\providecommand*{\mciteBstWouldAddEndPunctfalse}
  {\let\EndOfBibitem\relax}
\providecommand*{\mciteSetBstMidEndSepPunct}[3]{}
\providecommand*{\mciteSetBstSublistLabelBeginEnd}[3]{}
\providecommand*{\EndOfBibitem}{}
\mciteSetBstSublistMode{f}
\mciteSetBstMaxWidthForm{subitem}
{(\emph{\alph{mcitesubitemcount}})}
\mciteSetBstSublistLabelBeginEnd{\mcitemaxwidthsubitemform\space}
{\relax}{\relax}

\bibitem[Sun \emph{et~al.}(2018)Sun, Berkelbach, Blunt, Booth, Guo, Li, Liu,
  McClain, Sayfutyarova, Sharma, Wouters, and Chan]{Sun2018}
Q.~Sun, T.~C. Berkelbach, N.~S. Blunt, G.~H. Booth, S.~Guo, Z.~Li, J.~Liu,
  J.~D. McClain, E.~R. Sayfutyarova, S.~Sharma, S.~Wouters and G.~K.~L. Chan,
  \emph{WIREs Computational Molecular Science}, 2018, \textbf{8}, \relax
\mciteBstWouldAddEndPuncttrue
\mciteSetBstMidEndSepPunct{\mcitedefaultmidpunct}
{\mcitedefaultendpunct}{\mcitedefaultseppunct}\relax
\EndOfBibitem
\bibitem[Sun(2015)]{Libcint}
Q.~Sun, \emph{Journal of Computational Chemistry}, 2015, \textbf{36},
  1664--1671\relax
\mciteBstWouldAddEndPuncttrue
\mciteSetBstMidEndSepPunct{\mcitedefaultmidpunct}
{\mcitedefaultendpunct}{\mcitedefaultseppunct}\relax
\EndOfBibitem
\bibitem[McClean \emph{et~al.}(2020)McClean, Faulstich, Zhu, O'Gorman, Qiu,
  White, Babbush, and Lin]{McClean2020}
J.~R. McClean, F.~M. Faulstich, Q.~Zhu, B.~O'Gorman, Y.~Qiu, S.~R. White,
  R.~Babbush and L.~Lin, \emph{New Journal of Physics}, 2020, \textbf{22},
  093015\relax
\mciteBstWouldAddEndPuncttrue
\mciteSetBstMidEndSepPunct{\mcitedefaultmidpunct}
{\mcitedefaultendpunct}{\mcitedefaultseppunct}\relax
\EndOfBibitem
\bibitem[Sokolov \emph{et~al.}(2020)Sokolov, Barkoutsos, Ollitrault, Greenberg,
  Rice, Pistoia, and Tavernelli]{Sokolov2020}
I.~O. Sokolov, P.~K. Barkoutsos, P.~J. Ollitrault, D.~Greenberg, J.~Rice,
  M.~Pistoia and I.~Tavernelli, \emph{The Journal of Chemical Physics}, 2020,
  \textbf{152}, 124107\relax
\mciteBstWouldAddEndPuncttrue
\mciteSetBstMidEndSepPunct{\mcitedefaultmidpunct}
{\mcitedefaultendpunct}{\mcitedefaultseppunct}\relax
\EndOfBibitem
\bibitem[Virtanen \emph{et~al.}(2020)Virtanen, Gommers, Oliphant, Haberland,
  Reddy, Cournapeau, Burovski, Peterson, Weckesser, Bright, van~der Walt,
  Brett, Wilson, Millman, Mayorov, Nelson, Jones, Kern, Larson, Carey, Polat,
  Feng, Moore, VanderPlas, Laxalde, Perktold, Cimrman, Henriksen, Quintero,
  Harris, Archibald, Ribeiro, Pedregosa, van Mulbregt, Vijaykumar, Bardelli,
  Rothberg, Hilboll, Kloeckner, Scopatz, Lee, Rokem, Woods, Fulton, Masson,
  H{\"{a}}ggstr{\"{o}}m, Fitzgerald, Nicholson, Hagen, Pasechnik, Olivetti,
  Martin, Wieser, Silva, Lenders, Wilhelm, Young, Price, Ingold, Allen, Lee,
  Audren, Probst, Dietrich, Silterra, Webber, Slavi{\v{c}}, Nothman, Buchner,
  Kulick, Sch{\"{o}}nberger, {de Miranda Cardoso}, Reimer, Harrington,
  Rodr{\'{i}}guez, Nunez-Iglesias, Kuczynski, Tritz, Thoma, Newville,
  K{\"{u}}mmerer, Bolingbroke, Tartre, Pak, Smith, Nowaczyk, Shebanov, Pavlyk,
  Brodtkorb, Lee, McGibbon, Feldbauer, Lewis, Tygier, Sievert, Vigna, Peterson,
  More, Pudlik, Oshima, Pingel, Robitaille, Spura, Jones, Cera, Leslie, Zito,
  Krauss, Upadhyay, Halchenko, and V{\'{a}}zquez-Baeza]{Virtanen2020}
P.~Virtanen, R.~Gommers, T.~E. Oliphant, M.~Haberland, T.~Reddy, D.~Cournapeau,
  E.~Burovski, P.~Peterson, W.~Weckesser, J.~Bright, S.~J. van~der Walt,
  M.~Brett, J.~Wilson, K.~J. Millman, N.~Mayorov, A.~R. Nelson, E.~Jones,
  R.~Kern, E.~Larson, C.~J. Carey, Ä.~Polat, Y.~Feng, E.~W. Moore,
  J.~VanderPlas, D.~Laxalde, J.~Perktold, R.~Cimrman, I.~Henriksen, E.~A.
  Quintero, C.~R. Harris, A.~M. Archibald, A.~H. Ribeiro, F.~Pedregosa, P.~van
  Mulbregt, A.~Vijaykumar, A.~P. Bardelli, A.~Rothberg, A.~Hilboll,
  A.~Kloeckner, A.~Scopatz, A.~Lee, A.~Rokem, C.~N. Woods, C.~Fulton,
  C.~Masson, C.~H{\"{a}}ggstr{\"{o}}m, C.~Fitzgerald, D.~A. Nicholson, D.~R.
  Hagen, D.~V. Pasechnik, E.~Olivetti, E.~Martin, E.~Wieser, F.~Silva,
  F.~Lenders, F.~Wilhelm, G.~Young, G.~A. Price, G.~L. Ingold, G.~E. Allen,
  G.~R. Lee, H.~Audren, I.~Probst, J.~P. Dietrich, J.~Silterra, J.~T. Webber,
  J.~Slavi{\v{c}}, J.~Nothman, J.~Buchner, J.~Kulick, J.~L. Sch{\"{o}}nberger,
  J.~V. {de Miranda Cardoso}, J.~Reimer, J.~Harrington, J.~L.~C.
  Rodr{\'{i}}guez, J.~Nunez-Iglesias, J.~Kuczynski, K.~Tritz, M.~Thoma,
  M.~Newville, M.~K{\"{u}}mmerer, M.~Bolingbroke, M.~Tartre, M.~Pak, N.~J.
  Smith, N.~Nowaczyk, N.~Shebanov, O.~Pavlyk, P.~A. Brodtkorb, P.~Lee, R.~T.
  McGibbon, R.~Feldbauer, S.~Lewis, S.~Tygier, S.~Sievert, S.~Vigna,
  S.~Peterson, S.~More, T.~Pudlik, T.~Oshima, T.~J. Pingel, T.~P. Robitaille,
  T.~Spura, T.~R. Jones, T.~Cera, T.~Leslie, T.~Zito, T.~Krauss, U.~Upadhyay,
  Y.~O. Halchenko and Y.~V{\'{a}}zquez-Baeza, \emph{Nature Methods}, 2020,
  \textbf{17}, 261--272\relax
\mciteBstWouldAddEndPuncttrue
\mciteSetBstMidEndSepPunct{\mcitedefaultmidpunct}
{\mcitedefaultendpunct}{\mcitedefaultseppunct}\relax
\EndOfBibitem
\bibitem[Suzuki \emph{et~al.}(2020)Suzuki, Kawase, Masumura, Hiraga, Nakadai,
  Chen, Nakanishi, Mitarai, Imai, Tamiya, Yamamoto, Yan, Kawakubo, Nakagawa,
  Ibe, Zhang, Yamashita, Yoshimura, Hayashi, and Fujii]{suzuki2020qulacs}
Y.~Suzuki, Y.~Kawase, Y.~Masumura, Y.~Hiraga, M.~Nakadai, J.~Chen, K.~M.
  Nakanishi, K.~Mitarai, R.~Imai, S.~Tamiya, T.~Yamamoto, T.~Yan, T.~Kawakubo,
  Y.~O. Nakagawa, Y.~Ibe, Y.~Zhang, H.~Yamashita, H.~Yoshimura, A.~Hayashi and
  K.~Fujii, \emph{Qulacs: a fast and versatile quantum circuit simulator for
  research purpose}, 2020\relax
\mciteBstWouldAddEndPuncttrue
\mciteSetBstMidEndSepPunct{\mcitedefaultmidpunct}
{\mcitedefaultendpunct}{\mcitedefaultseppunct}\relax
\EndOfBibitem
\bibitem[Cowtan \emph{et~al.}(2020)Cowtan, Dilkes, Duncan, Simmons, and
  Sivarajah]{Cowtan2019}
A.~Cowtan, S.~Dilkes, R.~Duncan, W.~Simmons and S.~Sivarajah, \emph{Electronic
  Proceedings in Theoretical Computer Science}, 2020, \textbf{318},
  214--229\relax
\mciteBstWouldAddEndPuncttrue
\mciteSetBstMidEndSepPunct{\mcitedefaultmidpunct}
{\mcitedefaultendpunct}{\mcitedefaultseppunct}\relax
\EndOfBibitem
\bibitem[Grimsley \emph{et~al.}(2019)Grimsley, Economou, Barnes, and
  Mayhall]{Grimsley2019}
H.~R. Grimsley, S.~E. Economou, E.~Barnes and N.~J. Mayhall, \emph{Nature
  Communications}, 2019, \textbf{10}, 1--11\relax
\mciteBstWouldAddEndPuncttrue
\mciteSetBstMidEndSepPunct{\mcitedefaultmidpunct}
{\mcitedefaultendpunct}{\mcitedefaultseppunct}\relax
\EndOfBibitem
\end{mcitethebibliography}
\bibliographystyle{rsc}

\end{document}

\preprint{APS/123-QED}

\title{Supplementary Material: Molecular Excited State Calculations with Adaptive Wavefunctions on a Quantum Eigensolver Emulation
}

\author{Hans Hon Sang Chan}
 \email{hans.chan@materials.ox.ac.uk}
\affiliation{%
 Department of Materials, University of Oxford, Parks Road, Oxford OX1 3PH, United Kingdom
}%

\author{Nathan Fitzpatrick}%
\affiliation{%
 Cambridge Quantum Computing Ltd., 9a Bridge Street, Cambridge CB2 1UB, United Kingdom
}%

\author{Javier Segarra-Mart\'i}
\affiliation{%
 Instituto de Ciencia Molecular, Universitat de Valencia, PO Box 22085 Valencia, Spain
 }%

\author{Michael J. Bearpark}%
\affiliation{%
 Department of Chemistry, Molecular Sciences Research Hub, Imperial College London, White City Campus, 82 Wood Lane, London W12 0BZ , United Kingdom
}%

\author{David P. Tew}%
\affiliation{%
 Physical and Theoretical Chemical Laboratory, University of Oxford, South Parks Road, Oxford OX1 3QZ, United Kingdom
}%

\date{\today}% It is always \today, today,

\maketitle

The is a technical document detailing the functionalities of an open-sourced package, the \href{https://github.com/hanschanhs/QEBAB}{Quantum Eigensolver Building on Achievments of Both quantum computing and quantum chemistry (QEBAB)}, which we developed for investigating adaptive ansatz generation in excited state calculations in the Variational Quantum Eigensolver (VQE) framework.
The code interfaces a number of other open-sourced packages for quantum chemistry and quantum computing; PySCF\cite{Sun2018} and Libcint \cite{Libcint} for extracting the required one- and two-electron integrals and initialising the molecule, Open-Fermion\cite{McClean2020} for generation and transformation of UCC excitation operators, Pytket\cite{Sokolov2020} for construction and compilation of ansatz circuits, and any backend supported by Pytket for circuit simulation.
Additionally, in the associated work Scipy\cite{Virtanen2020} was employed for variational minimisation of ansatz expectation energies, as well as the computation of operator gradients for adaptive ansatz growth.
This document serves as a how-to guide to its usage (with annotated pseudo-code), and also as a step-by-step introduction to the VQE.

Typical workflow as follows: the user first initialises a molecule.
The package is used to generate a set of Unitary Coupled Cluster (UCC)-type excitation operators, and construct an ansatz wavefunction out of the operators according to the users choice.
The ansatz wavefunction is then transformed into a parameterised quantum circuit, which can be simulated on a number of quantum circuit simulator backends (in this work Qulacs \cite{suzuki2020qulacs} with GPU acceleration was used).
A classical optimiser is used to adjust the parameters in the quantum circuit until the simulated energy expectation value converges.

\section{Molecule Initialisation}
    Molecules are initialised as \verb|MolecularData| objects from OpenFermion:
    \begin{lstlisting}[language=Python]
    from openfermion import MolecularData

    # Initialise LiH with bond length 1.2 A
    molecule = MolecularData(geometry=[('Li', (0., 0., 0.)), ('H', (0., 0., 1.2))],
                    basis='sto-3g',
                    multiplicity=1)
    \end{lstlisting}
    With the OpenFermion PySCF wrapper, 1- and 2-electron integrals are calculated, which are stored in the \verb|MolecularData| object.
    In this work, Restricted Hartree-Fock (RHF) Self-Consistent Field (SCF) calculations were used.
    It is then possible to generate the second quantised molecular Hamiltonian from \verb|MolecularData|:
    \begin{equation}
    \hat{H} = h_\text{Nu} + \sum_{p,q}^N h_{pq}a^\dagger_p a_q + \frac{1}{2}\sum_{p,q,r,s}^N h_{pqrs}a^\dagger_p a^\dagger_q a_r a_s
    \end{equation}
    This needs to be modified into operations on a quantum computer.
    Using the Jordan-Wigner transform, the creation and annihilation operators ($a^\dagger_j$ and $a_j$) in the second quantised Hamiltonian from can be expressed in terms of Pauli matrices $\sigma_i \in\{\sigma^x, \sigma^y, \sigma^z\}$, which conveniently translate directly to 1-qubit Pauli logic gates $P_i \in \{X, Y, Z\}$:
    \begin{align}
        a_j^{\dagger} =&  \bigotimes^{j-1}_i \sigma_i^z \bigotimes \frac{1}{2}(\sigma^x_j - \text{i}\sigma^y_j) \label{eq:creation} \\
        =& \frac{1}{2}(X-\text{i}Y) \otimes [Z_{j-1} \otimes \dots \otimes Z_0]\\
        a_j =& \bigotimes^{j-1}_i \sigma_i^z \bigotimes \frac{1}{2}(\sigma^x_j + \text{i}\sigma^y_j) \\ 
        =& \frac{1}{2}(X+\text{i}Y) \otimes [ Z_{j-1} \otimes \dots \otimes Z_0 ] \label{eq:annihilation}
    \end{align}
    The second quantised qubit Hamiltonian is thus a linear combination of Pauli terms $\bigotimes_iP_i$ (each term is a tensor multiplication).
    The following is an example of the terms in the series:
    \begin{eqnarray}
        \hat{H} &=& h_0I + h_1Z_0 + \dots + h_7X_0\otimes Z_1\otimes X_2 + \dots \nonumber \\ 
        & & + h_{13}Y_0\otimes Z_1\otimes Y_2\otimes Z_3 + \dots \label{eq:exham}\\
        &=& \sum_j h_j \bigotimes_iP^j_i \label{eq:pauliham}
    \end{eqnarray}
    where $h_j$ are the 1 and 2 electron integrals and $h_0$ the nuclear Hamiltonian contribution from the SCF calculation.
    
    We use the \verb|jordan_wigner| transform function native to OpenFermion to map the fermionic operators in the Hamiltonian into unitary Pauli matrices that can be applied as quantum logic gates onto qubits.
    We chose to store the qubit Hamiltonian as Pytket \verb|QubitPauliOperator| objects.
    \begin{lstlisting}[language=Python]
    from openfermionpyscf import run_pyscf
    from openfermion.transforms import jordan_wigner
    from pytket import QubitPauliOperator
    
    # Calculate the 1,2 electron integrals
    molecule = run_pyscf(molecule, run_scf=1)
    
    # Hamiltonian
    ham_qubit = jordan_wigner(mol.get_molecular_hamiltonian())
    ham_qubit.compress() # Now in QubitOperator form
    ham = QubitPauliOperator.from_OpenFermion(ham_qubit)
    \end{lstlisting}

\section{Generating Excitation Operators}
\begin{figure*}
        \centering
        \includegraphics[scale=0.18]{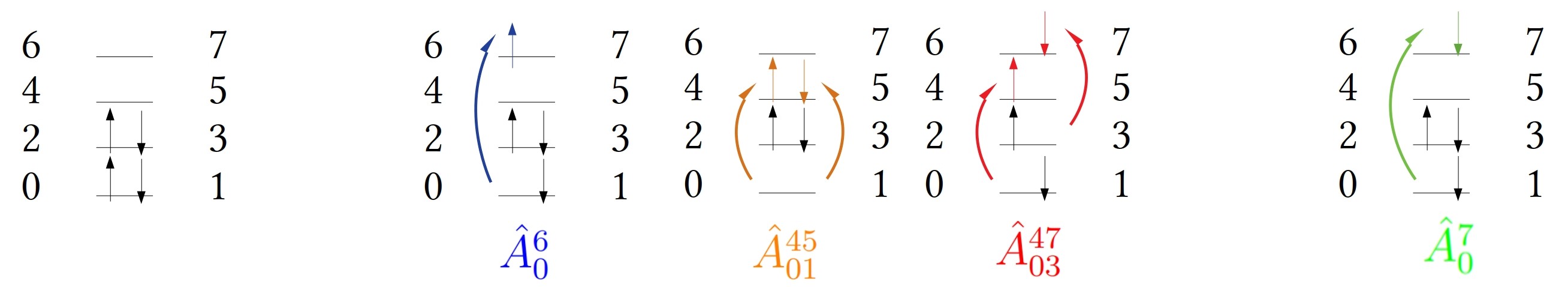}
        \caption{Single and double excitations. 
        Three diagrams in the middle express spin-preserving, physical excitations.
        Diagram in the far right does not preserve the spin, and is not generated.}
        \label{fig:excitations}
    \end{figure*}

    We created custom \verb|OperatorPool| classes which initialises different groups of UCC fermionic excitation operators from a given input number of electrons and number of orbitals.
    In the reported work, only the singlet-restricted, generalised excitations in the \verb|sUCCGSD_Pool| and \verb|sUpCCGSD_Pool| classes were used.
    Figure \ref{fig:excitations} further elaborates on which excitations are included
    The excitations are constructed using OpenFermion.
    Both inherit from the \verb|OperatorPool| class, which has a number of built-in excitation operator functions described in the following sections.
    \begin{lstlisting}[language=Python]
    from qebab.operators import sUpCCGSD_Pool
    
    pool = sUpCCGSD_Pool()
    pool.init(n_orb=molecule.n_orbitals,
              n_occ=molecule.get_n_alpha_electrons(),
              n_vir=molecule.n_orbitals - molecule.get_n_alpha_electrons())
    \end{lstlisting}
    
    \subsection{Pauli Gadgets}
        The \verb|OperatorPool| class transforms the fermionic excitation operators into strings of Pauli operators which maps onto quantum circuit components.
        Consider the parameterised UCC state preparation operator:
        \begin{equation}
            U(\vec{\theta}) = \prod_m e^{\theta_m(\tau_m - \tau_m^\dagger)} \label{eq:tUCC}
        \end{equation} 
        where $m$ indexes all possible single and double excitations, $\theta_m \in \{\theta^a_i, \theta^{ab}_{ij}\}$ and $\tau_m \in \{a^\dagger_a a_i, a^\dagger_a a^\dagger_b a_ia_j\}$.
        Using the Jordan-Wigner transform, each of the excitation terms $\theta_m(\tau_m-\tau^\dagger_m)$ are translated to Pauli matrices:
        \begin{equation}
        \theta_i^a(a^\dagger_i a_a - a^\dagger_a a_i) = \frac{\text{i}\theta^a_i}{2} \bigotimes_{k=i+1}^{a-1} \sigma^z_k(\sigma^y_i\sigma^x_a - \sigma^x_i\sigma^y_a)
        \end{equation}
        \begin{equation}
        \begin{split}
            \theta_{ij}^{ab} (a_i^\dagger a_j^\dagger a_aa_b - a_a^\dagger a_b^\dagger a_ia_j) =& \frac{\text{i}\theta_{ij}^{ab}}{8}  \bigotimes_{k=i+1}^{j-1} \sigma^z_k  \bigotimes_{l=a+1}^{b-1} \sigma^z_l \\
        &(\sigma^x_i \sigma^x_j \sigma^y_a \sigma^x_b + \sigma^y_i \sigma^x_j \sigma^y_a \sigma^y_b \\
        &+\sigma^x_i \sigma^y_j \sigma^y_a \sigma^y_b + \sigma^x_i \sigma^x_j \sigma^x_a \sigma^y_b \\
        &- \sigma^y_i \sigma^x_j \sigma^x_a \sigma^x_b - \sigma^x_i \sigma^y_j \sigma^x_a \sigma^x_b\\
        &- \sigma^y_i \sigma^y_j \sigma^y_a \sigma^x_b - \sigma^y_i \sigma^y_j \sigma^x_a \sigma^y_b )\\
        \end{split}
        \end{equation}
        where each product is a tensor multiplication.
        An exponentiated excitation term $e^{\theta_m(\tau_m-\tau^\dagger_m)}$ forms a string of circuit blocks (\textit{Pauli gadget}), each with a single qubit rotation gate parameterised to $\theta_m$ of the excitation term\cite{Cowtan2019}.
        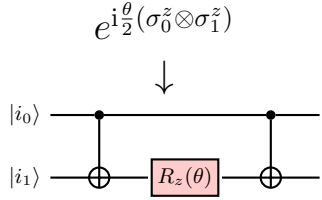
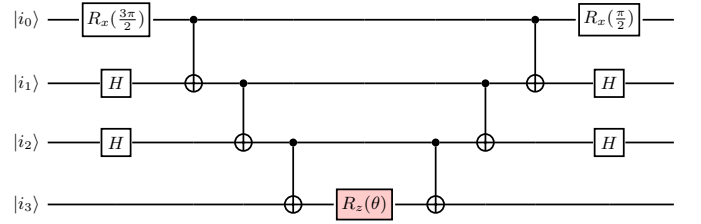
\begin{figure*}
            \centering
            \begin{subfigure}[c]{0.5\textwidth}
                \centering
                {\Large
                $e^{\text{i}\frac{\theta}{2}(\sigma^z_0 \otimes \sigma^z_1)} $\\
                $\downarrow$\\
                }
                \begin{quantikz}
                \lstick{$|i_0\rangle$}& \ctrl{1} & \qw & \ctrl{1} & \qw\\
                \lstick{$|i_1\rangle$}& \targ{} & \gate[style={fill=red!20}]{R_z(\theta)} & \targ{} &\qw
                \end{quantikz}
                \caption{
                    Creating the unitary gate $e^{\text{i}\frac{\theta}{2}(\sigma^z_0 \otimes \sigma^z_1)}$.
                    \textnormal{CNOT} gates are used to first entangle two qubits, then the rotation gate $R_z$ is applied, followed by a second \textnormal{CNOT} gate.
                }
            \end{subfigure}%
            ~
            \begin{subfigure}[c]{0.5\textwidth}
                \centering
                \resizebox{\textwidth}{!}{%
                \begin{quantikz}
                \lstick{$|i_0\rangle$} & \gate{R_x(\frac{3\pi}{2})} & \ctrl{1} & \qw & \qw &\qw & \qw & \qw & \ctrl{1} &\gate{R_x(\frac{\pi}{2})} & \qw \\
                \lstick{$|i_1\rangle$} & \gate{H} &\targ{}  & \ctrl{1} & \qw & \qw & \qw & \ctrl{1}  & \targ{} & \gate{H} & \qw \\
                \lstick{$|i_2\rangle$} & \gate{H} &\qw & \targ{}  & \ctrl{1} & \qw & \ctrl{1} & \targ{} & \qw & \gate{H} & \qw \\
                \lstick{$|i_3\rangle$} & \qw &\qw & \qw & \targ{}  & \gate[style={fill=red!20}]{R_z(\theta)} & \targ{}  & \qw & \qw & \qw & \qw \\
                \end{quantikz}
                }
                \caption{
                A Pauli gadget for $e^{\text{i}\frac{\theta}{2}(\sigma^y_0 \otimes \sigma^x_1 \otimes \sigma^x_2 \otimes \sigma^z_3)}$ in a 4 qubit circuit.
                The $R_x$ gates rotate the phase of a qubit; in the 0\textsuperscript{th} qubit it rotates the phase of the qubit to the $y$-basis.
                The Hadamard gates $H$ generate a superposition of the $\ket{0}$ and the $\ket{1}$ state; here it is used to access the $x$-basis of the qubit.
                }
            \end{subfigure}
            \caption{ Examples of Pauli gadgets.
            Note that the each excitation term is formed of multiple Pauli gadgets, all parameterised to the same $\theta$.
            }
        \end{figure*}

        Alternatively, the \verb|OperatorPool| class can also express excitation operators into unitary matrices that can be used for direct matrix evaluation of the expected circuit behaviour.

    \subsection{Analytic Gradients}
        The \verb|OperatorPool| class also computes the expected energy gradient of an ansatz (with respect to the free parameter $\theta_i$) for each candidate excitation operator $\hat{A}_i$ in the pool, a key aspect in the ADAPT methods.
        Consider first the derivative of a single qubit rotation gate. %(see Appendix \ref{appendix:gates}).
        We can represent the rotation matrices for the three cartesian axes as exponentials :
        \begin{align}
        R(\theta_i)_X =& e^{-\text{i}\theta_iX/2}\\
        R(\theta_i)_Y =& e^{-\text{i}\theta_iY/2}\\
        R(\theta_i)_Z =& e^{-\text{i}\theta_iZ/2}
        \end{align}
        Therefore for each differential with respect to the rotation angles there is only a single term in the sum:
        \begin{equation}
            \frac{\partial}{\partial \theta_i} R(\theta_i)_X = -\frac{\text{i}}{2} \cdot Xe^{-\text{i}\theta_iX/2}
        \end{equation}
        \begin{equation}
        \frac{\partial}{\partial \theta_i} R(\theta_i)_Y = -\frac{\text{i}}{2} \cdot Ye^{-\text{i}\theta_iY/2}
        \end{equation}
        \begin{equation}
            \frac{\partial}{\partial \theta_i} R(\theta_i)_Z = -\frac{\text{i}}{2} \cdot Ze^{-\text{i}\theta_iZ/2}
        \end{equation}
        Now consider an arbitrary unitary ansatz of the following form:
        \begin{equation}
            \ket{\Psi(\vec{\theta})} = U(\vec{\theta})\ket{\psi_\text{ref}}
        \end{equation}
        where the unitary is composed of exponentiated Pauli terms:
        \begin{equation}
            U(\vec{\theta}) = \prod^N_{i=1} e^{\theta_i\hat{A_i}}
        \end{equation}
        The energy expectation is:
        \begin{equation}
            E(\vec{\theta}) = \bra{\Psi(\vec{\theta})}\hat{H} \ket{\Psi(\vec{\theta})} 
        \end{equation}
        The energy gradient with respect to the $i^\text{th}$ parameter $\theta_i$ in the ansatz is thus:
        \begin{eqnarray}
            \pdv{E}{\theta_i} &=& \bra{\psi_\text{ref}} U^\dagger(\vec{\theta})\hat{H}\pdv{U(\vec{\theta})}{\theta_i} \ket{\psi_\text{ref}} \nonumber \\
            & & + \bra{\psi_\text{ref}} \pdv{U^\dagger(\vec{\theta})}{\theta_i} \hat{H}U(\vec{\theta}) \ket{\psi_\text{ref}}
        \end{eqnarray}
        where:
        \begin{equation}
            \pdv{U(\vec{\theta})}{\theta_i} = \prod^N_{j=i+1}(e^{\theta_j\hat{A}_j}) \hat{A_i}  \prod^i_{k=1}(e^{\theta_i\hat{A}_i})
        \end{equation}
        Substituting in:
        \begin{widetext}
        \begin{equation}
            \pdv{E}{\theta_i} = \bra{\Psi(\vec{\theta})}\hat{H} \prod^N_{j=i+1}(e^{\theta_j\hat{A}_j}) \hat{A_i}  \prod^i_{k=1}(e^{\theta_i\hat{A}_i}) \ket{\psi_\text{ref}} - \bra{\psi_\text{ref}} \prod^1_{k=i}(e^{-\theta_k\hat{A}_k})  \hat{A}_i \prod^{i+1}_{j=N}(e^{-\theta_j\hat{A}_j}) \hat{H}  \ket{\Psi(\vec{\theta})}
        \end{equation}
        \end{widetext}
        If we are only concerned with the last $m^\text{th}$ operator in the ansatz, this simplifies to:
        \begin{equation}
            \pdv{E}{\theta_m} = \bra{\Psi(\vec{\theta})}[\hat{H},\hat{A_m}]\ket{\Psi(\vec{\theta})}
        \end{equation}
        which is equivalent to:
        \begin{equation}
            \pdv{E}{\theta_m} = 2\mathcal{R}\bra{\Psi(\vec{\theta})}\hat{H}\hat{A_m}\ket{\Psi(\vec{\theta})}
        \end{equation}
        This is measured with the Hadamard test in Figure \ref{grad}.
        However our circuit is composed of Pauli gadgets rather than single qubit parameterized rotations such as in the UCC case.
        Following the same methodology using ancilla qubits and the Hadamard test, the circuit primitive corresponding to the derivative of a Pauli gadget with respect to its rotational parameter is given instead by Figure \ref{grad_pauli}.
        \begin{figure}[h!]
            \centering
            \begin{quantikz}
                && \lstick{$\ket{0}_a$} &  \gate{H} & \ctrl{1} & \gate{H} & \meter{} \\
                \lstick[wires=3]{$\ket{\psi}$}
                &&\lstick{$\ket{i}_0$} & \gate[wires=3, nwires=2][1cm]{U(\vec{\theta})} & \gate[wires=3, nwires=2]{V} & \qw & \qw \\
                && \vdots & & \\
                &&\lstick{$\ket{i}_N$} &\qw & & \qw & \qw 
            \end{quantikz}
            \caption{The Hadamard Test circuit for measuring the real part of the expected value when the unitary $V$ is applied to $\ket{\Psi(\vec{\theta})}$ i.e. $\textnormal{Re}\bra{\Psi(\vec{\theta})} V \ket{\Psi(\vec{\theta})} $.}
            \label{grad}
        \end{figure}
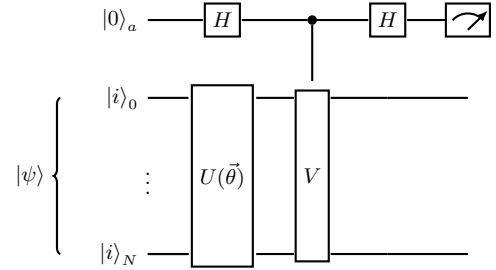
        \begin{figure*}
        \centering
        \begin{quantikz}
        \lstick{$|0\rangle$} & \gate{H}& \qw & \qw & \qw & \ctrl{4} & \qw & \qw & \qw&\qw & \gate{H}& \meter{}  \\
        \lstick{$|i_3\rangle$} & \gate{R_x(\frac{3\pi}{2})} & \ctrl{1} & \qw & \qw &\qw & \qw & \qw & \qw & \ctrl{1} &\gate{R_x(\frac{\pi}{2})} & \qw \\
        \lstick{$|i_2\rangle$} & \gate{H} &\targ{}  & \ctrl{1} & \qw & \qw &\qw &  \qw & \ctrl{1}  & \targ{} & \gate{H} & \qw \\
        \lstick{$|i_1\rangle$} & \gate{H} &\qw & \targ{}  & \ctrl{1} & \qw & \qw & \ctrl{1} & \targ{} & \qw & \gate{H} & \qw \\
        \lstick{$|i_0\rangle$} & \qw &\qw & \qw & \targ{} & \gate{Z}  & \gate{R_z(\theta)} & \targ{}  & \qw & \qw & \qw & \qw \\
        \end{quantikz}
        \caption{Circuit Primitive for $\mathcal{R}(\frac{\delta}{\delta \theta_i}e^{i\frac{\theta}{2}(\sigma^z_0 \otimes \sigma^x_1 \otimes \sigma^x_2 \otimes \sigma^y_3)})$ }\
        \label{grad_pauli}
        \end{figure*}
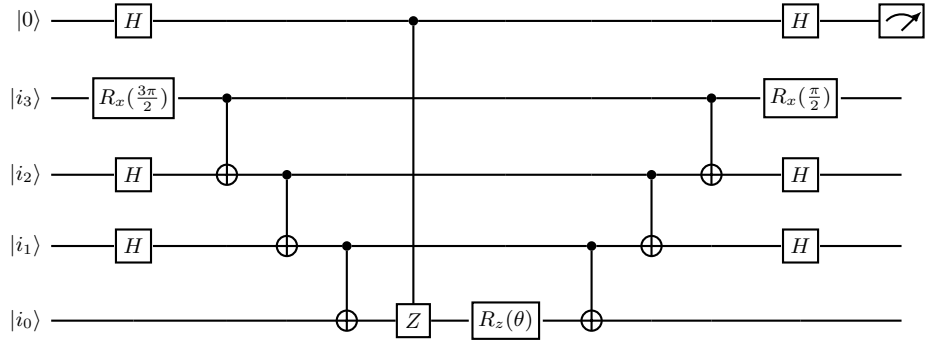
        
        For adaptively growing the ansatz for excited states, further gradients of the overlap with other eigenstates need to be measured:
        \begin{widetext}
        \begin{align}
        \pdv{E}{\theta_m} =& 2\mathcal{R} \bra*{\Psi(\vec{\theta})}\hat{A}_m\hat{H_k}\ket*{\Psi(\vec{\theta})} \label{eq:adapt-vqd}\\
        =& 2\mathcal{R} \bra*{\Psi(\vec{\theta})}\hat{A}_m  \left(\hat{H} + \sum^{j-1}_{i=0} \beta_i \ket*{\Phi_i}\bra*{\Phi_i} \right)  \ket*{\Psi(\vec{\theta})} \nonumber\\
        =& 2\mathcal{R} \bra*{\Psi(\vec{\theta})}\hat{A}_m\hat{H} + \hat{A}_m\sum^{j-1}_{i=0} \beta_i \ket*{\Phi_i}\bra*{\Phi_i} \ket*{\Psi(\vec{\theta})} \nonumber\\
        =& 2\mathcal{R} \bra*{\Psi(\vec{\theta})}\hat{A}_m\hat{H}\ket*{\Psi(\vec{\theta})} +  \underbrace{2\mathcal{R}\sum^{j-1}_{i=0} \beta_i\bra*{\Psi(\vec{\theta})}\hat{A}_m \ket*{\Phi_i}\bra*{\Phi_i}\ket*{\Psi(\vec{\theta})}}_\text{overlap gradient}
        \end{align}
        \end{widetext}
        Details of circuits for overlap measurement to follow.
        
        Inspired by the original creators of the ADAPT method\cite{Grimsley2019}, these gradient computations are computed by evaluating the corresponding unitary matricies instead of simulating circuits which would provide analytic gradients of operators.
        The code does not implement calculation of the gradient in circuit form.

\section{Reference Circuits and Ansatz Generation}
\begin{figure*}
            \centering
            \resizebox{\textwidth}{!}{%
            \begin{quantikz}
                &&\lstick{$\ket{0}_0$} & \gate{X} \slice{} & \gate{R_x(\frac{3\pi}{2})} & \ctrl{1} & \qw & \qw & \qw & \ctrl{1} & \gate{R_x(\frac{\pi}{2})}\slice{} & \gate{H} & \ctrl{1} & \qw & \qw & \qw & \ctrl{1} & \gate{H}\slice{}                                     & \gate{R_x(\frac{3\pi}{2})} & \ctrl{1} & \qw & \qw &\qw & \qw & \qw & \ctrl{1} &\gate{R_x(\frac{\pi}{2})} & \qw & \\
                &&\lstick{$\ket{0}_1$} & \gate{X}                      & \qw & \targ{} & \ctrl{1} & \qw & \ctrl{1} & \targ{} & \qw                                                          & \qw & \targ{} & \ctrl{1} & \qw & \ctrl{1} & \targ{} & \qw                                                         & \gate{R_x(\frac{3\pi}{2})} &\targ{}  & \ctrl{1} & \qw & \qw & \qw & \ctrl{1}  & \targ{} & \gate{R_x(\frac{\pi}{2})} & \qw &  \\
                &&\lstick{$\ket{0}_2$} & \qw                           & \gate{H} & \qw & \targ{} & \gate[style={fill=blue!20}]{\theta_1} & \targ{} & \qw & \gate{H}                         & \gate{R_x(\frac{3\pi}{2})} & \qw & \targ{} & \gate[style={fill=blue!20}]{\theta_1} & \targ{} & \qw & \gate{R_x(\frac{\pi}{2})} & \gate{R_x(\frac{3\pi}{2})} &\qw & \targ{}  & \ctrl{1} & \qw & \ctrl{1} & \targ{} & \qw & \gate{R_x(\frac{\pi}{2})} & \qw &  \\
                &&\lstick{$\ket{0}_3$} & \qw                           & \qw & \qw & \qw & \qw & \qw & \qw & \qw                                                                            & \qw & \qw & \qw & \qw & \qw & \qw & \qw                                                                 & \gate{H} &\qw & \qw & \targ{}  & \gate[style={fill=green!20}]{\theta_2} & \targ{}  & \qw & \qw & \gate{H} & \qw & 
            \end{quantikz}
            }%
            $$\hspace{0.2cm}\underbrace{\hspace{1.3cm}} \hspace{0.2cm} \underbrace{\hspace{9.5cm}} \hspace{0.5cm} \underbrace{\hspace{6.4cm}}$$ 
            {\Large
            \begin{equation*}
                \hspace{0.3cm} \ket{\psi_\text{HF}} \hspace{4cm} e^{\textcolor{blue}{\theta_1}(\tau_1-\tau_1^\dagger)} \hspace{6.2cm} e^{\textcolor{green}{\theta_2}(\tau_2-\tau_2^\dagger) \qquad \dots \hspace{3cm}} 
            \end{equation*}
            }%
            \caption{Section of a UCC-type state preparation circuit $\ket{\Psi(\vec{\theta})} = \dots e^{\theta_2(\tau_2-\tau_2^\dagger)}e^{\theta_1(\tau_1-\tau_1^\dagger)}\ket{\psi_\textnormal{HF}}$.
            }
        \label{fig:state_circ}
    \end{figure*}
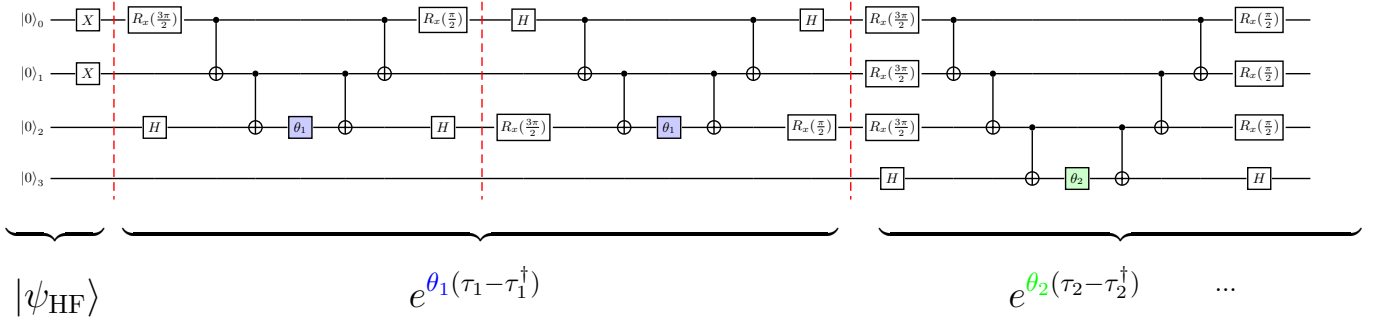

    We created custom \verb|Ansatz| constructor classes which take a pool of excitation operators in \verb|OperatorPool| and a reference circuit as input, then creates a symbolically parameterised state preparation ansatz circuit of choice.
    A number of different references have been defined, and in this work the closed-shell singlet HF reference $\ket{\psi_\text{HF}}$ and open-shell lowest energy triplet reference $\ket{\psi_{T_1}}$ were used (see main text). 
    We use the Pytket circuit generator to build the ansatz circuits from sequences of Pauli operators, and also Pytket compilation passes to reduce the quantum gate count in the circuit.
    Figure \ref{fig:state_circ} is an example of a constructed circuit.
    In this work, the \verb|k_UCC_Ansatz| and the \verb|ADAPT_VQD_Ansatz| constructors were used.
    
    The \verb|k_UCC_Ansatz| can be used to repeat a set of excitation operators $k$ times, as prescribed in the $k$-UpCCGSD ansatz.
    As it is a fixed ansatz for every eigenstate and geometry, it only needs to be called once in the beginning of a calculation.
    \begin{lstlisting}[language=Python]
    class k_UCC_Ansatz(Ansatz):
        def generate_Circuit(self, ref: str, k: int):
            ...
            # Reference Circuit
            ref_circ = reference_circuit_lib[ref]
            ...
            # Ansatz building, k-depth
            self.symbols = {}
            for rep in range(1, k+1):
                qubit_pauliop = {}
                for i in range(n_params):
                    # Generate fresh symbol
                    theta = fresh_symbol('t{}'.format(i))
                    self.symbols[theta] = None
                    # Isolate operator
                    op = self.pool.qubit_paulistrs[i]
                    for qpstr, coeff in op.items():
                        if coeff.imag > 0:
                            qubit_pauliop[qpstr] = theta
                        else:
                            qubit_pauliop[qpstr] = -1.0 * theta
                
                Pauli_U = QubitPauliOperator(qubit_pauliop)
    
                if rep==1:
                    sym_circ = gen_term_sequence_circuit(Pauli_U,
                                                         ref_circ,
                                                         partition_strat=PauliPartitionStrat.CommutingSets,
                                                         colour_method=GraphColourMethod.Lazy)
                else:
                    k_circ = Circuit(n_qubits)
                    k_circ = gen_term_sequence_circuit(Pauli_U,
                                                       k_circ,
                                                       partition_strat=PauliPartitionStrat.CommutingSets,
                                                       colour_method=GraphColourMethod.Lazy)
    
                    sym_circ.append(k_circ)
            ...
            # Compilation pass
            self.smart_circ = sym_circ.copy()
            Transform.UCCSynthesis(PauliSynthStrat.Sets, CXConfigType.Tree).apply(self.smart_circ)
            
            return self.smart_circ, self.symbols
    \end{lstlisting}
    
    \begin{figure*}
        \centering
        \resizebox{0.87\textwidth}{!}{%
        \begin{subfigure}[c]{0.5\textwidth}
            \centering
            \begin{equation*}
                \qquad \hat{H} = h_0I + \dots + h_j\textcolor{red}{\underline{\underline{ \bigotimes_i P^j_i}}} + h_k\bigotimes_i P^k_i +  \dots
            \end{equation*}
            $$\quad\qquad \Downarrow$$
            \begin{quantikz}
                \lstick[wires=4]{$\ket{\psi}$}
                &&\lstick{$\ket{i_0}$} &\qw& \gate[wires=4,nwires={3},style={fill=red!20}][2cm]{\bigotimes_i P_i} & \qw & \meter{}  \\
                &&\lstick{$\ket{i_1}$} &\qw & & \qw & \meter{}  \\
                & & \lstick{$\vdots$} & & & & \lstick{$\vdots$}  \\
                &&\lstick{$\ket{i_n}$} & \qw & & \qw & \meter{}  \\
            \end{quantikz}
        \end{subfigure}%
        \begin{subfigure}[c]{0.5\textwidth}
            \centering
            \begin{tabular}{ccc}
                \toprule
                Outcome & Parity & Eigenvalue \\
                \colrule
                $\dots000$ & even & +1 \\
                $\dots001$ & odd & -1 \\
                $\dots101$ & even & +1 \\
                $\dots111$ & odd & -1 \\
                $\vdots$ & $\vdots$ & $\vdots$ \\
                \colrule
              \end{tabular}
              \begin{equation*}
                \bra{\psi} \bigotimes_i P_i^j\ket{\psi} = \frac{\sum \text{Measured Eigenvalues}}{\text{Number of Measurements}}
              \end{equation*}
        \end{subfigure}
        }
        \caption{
        (LEFT) Circuit for measuring one Pauli term.
        Upon measuring, each qubit collapses to either 0 or 1.
        The is repeated multiple times and the results (shots) are recorded.
        (RIGHT) A shot table mapping each measurement outcome to an eigenvalue (parity is whether the number of 1's in the measurement is odd or even, which corresponds to an eigenvalue of -1 and +1 respectively).
        The energy expectation of the Pauli term $P_i$ is thus the average eigenvalue over multiple shots multiplied by $h_i$.
        }
        \label{fig:meas}
    \end{figure*}
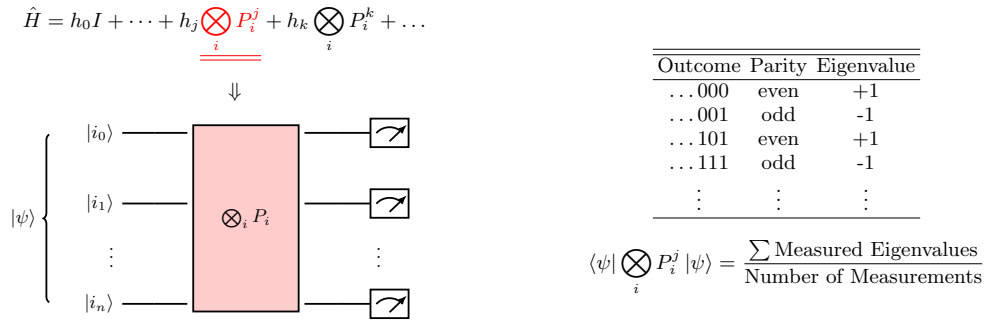
    
    The \verb|ADAPT_VQD_Ansatz| is of course adaptive and needs to be called at each new geometry.
    The constructor will iteratively grow an ansatz until the convergence criterion is met, and so requires the convergence threshold $\epsilon$ as input.
    It also needs to compute the energy and overlap gradient.
    For the latter, it calls the analytic gradient functions described above from the \verb|Operator_Pool| classes.
    \begin{lstlisting}[language=Python]
    class ADAPT_VQD_Ansatz(Ansatz):
        def generate_Circuit(self,
                            ref: str,
                            params: list, # currently sought after state
                            eigen_ansatze: list, # list of circuits
                            beta: float,
                            ham_sparse,
                            threshold):
                    
            if len(params)==0: # new eigenstate
                # reset for new eigenstate
                self.smart_circ = None 
                self.symbols = {}
                self.f_op = []
                
                # reset Reference Circuit
                ref_circ = reference_circuit_lib[ref]
                
                qubit_pauliop = {}
                Pauli_U = QubitPauliOperator(qubit_pauliop)
                self.grad_circ = gen_term_sequence_circuit(Pauli_U,
                                                           self.ref_circ,
                                                           partition_strat=PauliPartitionStrat.CommutingSets,
                                                           colour_method=GraphColourMethod.Lazy)
                self.smart_circ = self.grad_circ.copy()
    
            else: # repopulate current eigenstate
                self.grad_circ = self.smart_circ.copy()
                self.symbols = dict(zip(self.symbols, params))
                self.grad_circ.symbol_substitution(self.symbols)
    
    
            # Calculating gradients for operators in pool
            curr_norm = 0
            next_deriv = 0
            for op_index in range(self.pool.n_ops):
                ...    
                # Energy Gradient
                gi = self.pool.compute_gradient_i(op_index,
                                             ham_sparse,
                                             self.grad_circ,
                                             backend)
                # Overlap Gradient
                overlap_list = []
                for eigen_circ in eigen_ansatze:
                    # 2 Re beta * <ansatz|A(k)|eigen><eigen|ansatz>
                    overlap = sqrt(gen_overlap(self.grad_circ, eigen_circ, backend))
                    ov_g = abs(self.pool.compute_ov_grad_i(op_index, self.grad_circ, eigen_circ, backend))
                    overlap_list.append(abs(np.real(2 * beta * ov_g * overlap)))
                overlap_sum = sum(overlap_list)
    
                ...
                # Add up total gradient of operator
                gi = abs(gi) + overlap_sum
    
                curr_norm += gi**2
                if abs(gi) > next_deriv:
                    next_deriv = abs(gi)
                    next_index = op_index
    
            curr_norm = np.sqrt(curr_norm)
            max_of_com = next_deriv
    
            # Convergence or growth
            if curr_norm < threshold:
                self.converged = True
            else:
                qubit_pauliop = {}
                op_circ = Circuit(self.pool.n_spin_orb)
    
                # Generate fresh symbol
                theta = fresh_symbol('t')
                self.symbols[theta] = None
    
                # Append fermion operator
                self.f_op.append(self.pool.fermi_ops[next_index])
                self.op_indices.append(next_index)
    
                # Isolate operator
                op = self.pool.qubit_paulistrs[next_index]
                for qpstr, coeff in op.items():
                    if coeff.imag > 0:
                        qubit_pauliop[qpstr] = theta
                    else:
                        qubit_pauliop[qpstr] = -1.0 * theta
    
                Pauli_U = QubitPauliOperator(qubit_pauliop)
                op_circ = gen_term_sequence_circuit(Pauli_U,
                                                    op_circ,
                                                    partition_strat=PauliPartitionStrat.CommutingSets,
                                                    colour_method=GraphColourMethod.Lazy)
                self.smart_circ.append(op_circ)
                ...    
                Transform.UCCSynthesis(PauliSynthStrat.Sets, CXConfigType.Tree).apply(self.smart_circ)
                
            return self.converged, self.smart_circ, self.symbols, self.final_map
    \end{lstlisting}

\section{Ansatz Optimisation and Energy Calculation}
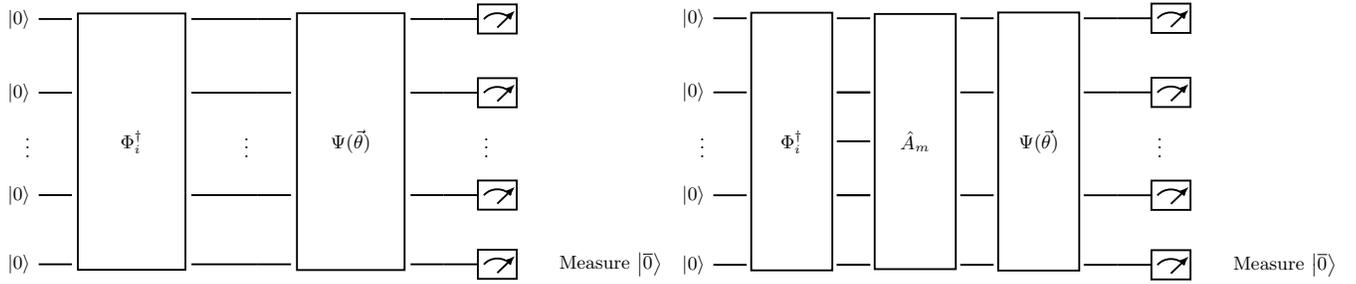
\begin{figure*}
        \centering
        \begin{subfigure}[c]{0.5\textwidth}
            \centering
            \resizebox{\textwidth}{!}{%
            \begin{quantikz}
            \lstick{$\ket{0}$}& \gate[wires=5,nwires={3}][2cm]{\Phi_i^\dagger} & [0.5cm] \qw & \gate[wires=5, nwires={3}][2cm]{\Psi(\vec{\theta})} & \qw & \meter{}  \\
            \lstick{$\ket{0}$} & & \qw & & \qw & \meter{}  \\
            \lstick{$\vdots$} & & \lstick{$\vdots$} & & & \lstick{$\vdots$}   \\
            \lstick{$\ket{0}$} & & \qw & & \qw & \meter{}  \\
            \lstick{$\ket{0}$} & & \qw & & \qw & \meter{} & \rstick{Measure $\ket{\overline{0}} $}  \\
            \end{quantikz}
            }
        \end{subfigure}%
        \begin{subfigure}[c]{0.5\textwidth}
            \centering
            \resizebox{\textwidth}{!}{%
            \begin{quantikz}
            \lstick{$\ket{0}$}& \gate[wires=5,nwires={3}][1.5cm]{\Phi_i^\dagger} & \gate[wires=5,nwires={3}][1.5cm]{\hat{A}_m} & \gate[wires=5, nwires={3}][1.5cm]{\Psi(\vec{\theta})} & \qw & \meter{}  \\
            \lstick{$\ket{0}$} & & \qw & & \qw & \meter{}  \\
            \lstick{$\vdots$} & & \qw & & & \lstick{$\vdots$}   \\
            \lstick{$\ket{0}$} & & \qw & & \qw & \meter{}  \\
            \lstick{$\ket{0}$} & & \qw & & \qw & \meter{} & \rstick{Measure $\ket{\overline{0}} $}\\
            \end{quantikz}
            }
        \end{subfigure}
        \caption{(LEFT) the vacuum test, which doubles the circuit depth.
        (RIGHT) Measurement of the overlap gradient. 
        }
        \label{fig:overlapcirc}
    \end{figure*}

    The total energy expectation is the sum of energy expectation of each Pauli terms in the qubit Hamiltonian:
    \begin{align}
        \langle E\rangle =& \bra{\Psi(\vec{\theta})}\hat{H}\ket{\Psi(\vec{\theta})} \\
        =& \sum_j h_j \bra{\Psi(\vec{\theta})} \bigotimes_i P_i^j\ket{\Psi(\vec{\theta})}
    \end{align}
    Figure \ref{fig:meas} elaborates on the statistical nature of these measurements.
    In excited state VQD calculations, the overlap is calculated using the vacuum test:
    \begin{equation*}
        \textnormal{since}\quad  \ket{\Phi_i} =V_i\ket{0\dots00}  \quad \text{ and } \quad \ket{\Psi(\vec{\theta})} = U\ket{0\dots00} 
    \end{equation*}
    where $V_i$ and $U$ are unitary reference and state preparation operators, then the overlap between the two states is the expectation of measuring $\ket{0\dots00}$ when the $V^\dagger_iU$ circuit is applied.
    \begin{equation*}
        \bra{\Phi_i}\ket{\Psi(\vec{\theta})} = \bra{0\dots00}V^\dagger_iU\ket{0\dots00}
    \end{equation*}
    This technique doubles the depth of the circuit (refer to Figure \ref{fig:overlapcirc} for the circuit diagram).
    \begin{lstlisting}[language=Python]
    def get_zero_state_probability(circ):
        """Measures qubits in 0 basis
        """
        statevector = ProjectQBackend().get_state(circ)

        return abs(statevector[0])**2


    def overlap_gen(psi_circ, phi_circ):
        """Overlap measurement: vacuum test
        """
        circ = psi_circ.copy()
        phi_circ = phi_circ.dagger()
        circ.append(phi_circ)

        Transform.OptimisePhaseGadgets().apply(circ)
        prob_X = get_zero_state_probability(circ=circ)

        return prob_X
    \end{lstlisting}

    Combining the above components, an objective function which calculates expectation values (with orthogonal penalisation included for excited states) given an input of free parameters \verb|thetas| is needed.
    Any backend supported by Pytket can be used to simulate this measurement; in this investigation expectation calculations were performed on the noiseless quantum simulator ProjectQ.
    \begin{lstlisting}[language=Python]
    def gen_objective(operator_pool, qubit_ham, optimised_ansatze, beta):
        def energy_objective(thetas):
            """(pseudo-code) example of energy objective function
                which also accounts for the overlap
            """
            circ = gen_ansatz_circ(thetas, operator_pool)
            # E = <psi|H|psi>
            energy = ProjectQBackend().get_expectation_value(circ,qubit_ham)
            
            overlap_sums = []
            if len(optimised_ansatze)!=0:
                for phi_i in optimised_ansatze:
                    # b * <phi_i|psi>
                    overlap_i = beta * overlap_gen(circ, phi_i)
                    overlap_sums.append(overlap_i)

            overlap_sums = sum(overlap_sums)
            energy = energy + overlap_sums

            return energy
        return energy_objective
    \end{lstlisting}

    This objective function was used for the nonlinear classical optimisation of the wavefunction ansatz and energy eigenvalues with respect to the free parameters.
    This was achieved with the iterative \verb|optimize.minimize| function from Scipy using the Limited Broyden-Fletcher-Goldfarb-Shannon ``Bound" (L-BFGS-B) method.
    The L-BFGS-B used norm of the projected energy gradient smaller than $10^{-5}$ as convergence criterion, with a maximum number of iteration set at 30.
    The initial input values for the free parameters were chosen to be random numbers distributed between 0 and 0.1 throughout.

\section{Spin Expectation Calculation}
    In this investigation the $\hat{S}^2$ expectation values of optimised ans\"atze were computed to verify spin-restrictions were observed. 
    The $\hat{S}^2$ operator is:
    \begin{equation}
        \hat{S}^2 = \hat{S}_+\hat{S}_- + \hat{S}_z(\hat{S}_z-1)
    \end{equation}
    where:
    \begin{align}
        \hat{S}_+ =& \sum_p a^\dagger_{p_\alpha}a_{p_\beta}\\
        \hat{S}_- =& \sum_q a^\dagger_{q_\beta}a_{q_\alpha} \\
        \hat{S}_z =& \frac{1}{2}\sum_p (a^\dagger_{p_\alpha}a_{p_\alpha} - a^\dagger_{p_\beta}a_{p_\beta})
    \end{align}
    The spin operator in second quantisation is thus:
    \begin{widetext}
    \begin{eqnarray}
        \hat{S}^2 &=& \sum_{p,q}a^\dagger_{p_\alpha}a_{p_\beta} a^\dagger_{p_\beta}a_{p_\alpha} + \frac{1}{2}\sum_p (a^\dagger_{p_\alpha}a_{p_\alpha} - a^\dagger_{p_\beta}a_{p_\beta}) \left[\frac{1}{2}\sum_q (a^\dagger_{p_\alpha}a_{p_\alpha} - a^\dagger_{p_\beta}a_{p_\beta}) -1\right] \nonumber \\
        &=& \sum_{p,q} \left[ a^\dagger_{p_\alpha}a_{p_\beta} a^\dagger_{p_\beta}a_{p_\alpha}
        + \frac{1}{4}\left(a^\dagger_{p_\alpha}a_{p_\alpha} a^\dagger_{q_\alpha}a_{q_\alpha} - a^\dagger_{p_\alpha}a_{p_\alpha} a^\dagger_{q_\beta}a_{q_\beta} - a^\dagger_{p_\beta}a_{p_\beta} a^\dagger_{q_\alpha}a_{q_\alpha} + a^\dagger_{p_\beta}a_{p_\beta} a^\dagger_{q_\beta}a_{q_\beta}\right) \right] \nonumber \\
        & & - \frac{1}{2} \sum_p (a^\dagger_{p_\alpha}a_{p_\alpha} - a^\dagger_{p_\beta}a_{p_\beta})
    \end{eqnarray}
    \end{widetext}
    \newpage
    The above operator is defined as a \verb|FermionOperator| object for a molecule with a given number of orbitals.
    Once a converged ansatz is obtained, its $\hat{S}^2$ expectation value is obtained using the same procedure as that of energy expectation calculation, substituting the Hamiltonian \verb|FermionOperator| with the $\hat{S}^2$ \verb|FermionOperator|.

\bibliography{library}
\bibliographystyle{rsc}